\documentclass[aps,prb,twocolumn,letterpaper,showpacs,10pt]{revtex4-1}
\usepackage{graphicx,amsmath,mathrsfs}
\usepackage{amssymb}
\usepackage{amsthm}
\usepackage{bm}
\usepackage{float}
\usepackage{epsfig}

\usepackage{color}

\def\bc{\begin{center}}
\def\ec{\end{center}}

\renewcommand {\ec}{\eta_{\gamma}}

\newcommand{\comment}[1]{}

\usepackage{color}

\begin{document}
\title{Generalized Exact Holographic Mapping with Wavelets}
\author{Ching Hua Lee}
\affiliation{Institute of High Performance Computing, 138632, Singapore}
\email{calvin-lee@ihpc.a-star.edu.sg}
\affiliation{Department of Physics, National University of Singapore,117542, Singapore}
\email{phylch@nus.edu.sg}

\date{\today}
\begin{abstract}
The idea of renormalization and scale invariance is pervasive across disciplines. It has not only drawn numerous surprising connections between physical systems under the guise of holographic duality, but has also inspired the development of wavelet theory now widely used in signal processing. Synergizing on these two developments, we describe in this paper a generalized exact holographic mapping that maps a generic N-dimensional lattice system to a N+1-dimensional holographic dual, with the emergent dimension representing scale. In previous works, this was achieved via the iterations of the simplest of all unitary mappings, the Haar mapping, which fails to preserve the form of most Hamiltonians. By taking advantage of the full generality of biorthogonal wavelets, our new generalized holographic mapping framework is able to preserve the form of a large class of lattice Hamiltonians. By explicitly separating features that are fundamentally associated with the physical system from those that are basis-specific, we also obtain a clearer understanding of how the resultant bulk geometry arises. For instance, the number of nonvanishing moments of the high pass wavelet filter is revealed to be proportional to the radius of the dual Anti deSitter (AdS) space geometry. We conclude by proposing modifications to the mapping for systems with generic Fermi pockets.



\end{abstract}
\maketitle

\section{Introduction}

The theme of holographic duality has fascinated a generation of physicists in both high energy and condensed matter circles. Also known as the Anti-de-Sitter space/Conformal Field Theory (AdS/CFT) correspondence, it was pioneered by Witten, Maldacena, Klebanov and others\cite{maldacena1998,witten1998,gubser1998,witten1998b} in 1998, when an equivalence was made between a $D+1$-dimensional quantum field theory a $D+2$-dimensional gravitational theory at the partition function level. The canonical example of holographic duality is the correspondence between 3+1-dimensional super-Yang-Mills theory and 4+1-dimensional supergravity, with the large $N$ (strongly-coupled) limit of the super-Yang-Mills theory being dual to the classical (weakly-coupled) limit of the gravitational theory. At the core of holographic duality is the interpretation of a quantum field theory as a "hologram" of a dual gravitational system with one higher dimension, with the extra emergent dimension representing scale. This provides an avenue to understanding renormalization group (RG) flow dynamics in terms of bulk gravitational dynamics\cite{akhmedov1998,freedman1999renormalization,boer2000,skenderis2002,heemskerk2011,lee2010,han2017loop}. Inspired by that, holographic duality has also been used as a tool for understanding the nature of 
quantum criticality and high temperature superconductivity\cite{hartnoll2009,horowitz2009,mcgreevy2010,sachdev2012}, for which the exact role of the underlying strong coupling mechanism remains elusive.  

In face of evidence for the existence of holographic duality in various contexts, it will be very desirable to have a microscopic description of holography. This allows for a clear, constructive approach for understanding the dual theory, when it exists. For this purpose, an approach known as the Exact Holographic Mapping (EHM) was proposed by Qi\cite{qi2013} for generic lattice systems. Through recursive applications of local unitary transforms, this mapping maps a given ``boundary'' system onto a ``bulk'' system with a unitary equivalent Hilbert space, but having an extra emergent dimension representing scale\cite{huang2015entanglement,singh2016holographic,de2016tensor}. Geodesics distances in the bulk system can be determined from the decay behavior of their correlators. Although bulk systems obtained in this way via the EHM are not semiclassical bulk geometries corresponding to the large $N$ limit, in the strict sense of AdS-CFT, they possess geometries agreeing with expectations from the Ryu-Takayanagi formula\cite{ryu2006}. Notable examples include the AdS bulk geometry from a critical boundary fermion at zero temperature, and the BTZ (Ba˜\~{n}ados, Teitelboim, and Zanelli)\cite{banados1992} black hole geometry at nonzero temperature. As shall be elaborated in this paper, these geometric properties arise due to the fundamental scaling behaviors of the systems under consideration, and holds even for $N=1$ free fermions. Besides defining a bulk geometry, the EHM procedure is also useful in analyzing the RG properties of topological quantities. For instance, the holographic decomposition of the Berry curvature of a boundary Chern insulator interestingly reveals a $\mathbb{Z}_2$ topological insulator living in the holographic bulk, thereby providing a holographic interpretation of the parity anomaly\cite{gulee2016}. 


Parallel to these developments in holography is the development of wavelet transforms in computer science, with applications ranging from image compression to multiscale music texture to financial data analysis. In essence, wavelet transforms are ``lossless'' RG transforms\footnote{Usually, renormalization group analysis involve integrating out small-scale degrees of freedom that are deemed irrelevant, thereby losing information.} probing details of different spatial or temporal scales, very analogous to the objective of holography. As such, there has been a symbiosis of ideas between these two developments; in fact, the EHM is mathematically a Haar wavelet transform acting on the quantum mechanical Hilbert space rather than the space of signals. Recently, wavelets bases have also been shown to provide good approximations\cite{evenbly2016entanglement,evenbly2016representation,matsueda2016analytic} to certain critical ground states in the framework of the multi-scale entanglement renormalization ansatz (MERA)\cite{white1992,klumper1993,verstraete2004,vidal2007,vidal2008,gu2009,swingle2012,swingle2012b,evenbly2011,nozaki2012,hartman2013,czech2014,miyaji2014,miyaji2015,pastawski2015,czech2017defect}, a tensor network approach pioneered by Vidal et.al. that is closely related to the EHM\cite{vidal2008class,aguado2008entanglement,konig2009exact}
. Described as a quantum circuit, the EHM has proposed implementations with Gaussian entangled states in optical networks, circuit QED setups as well as trapped cold ions\cite{singh2016holographic,wang2014weaving,chen2014experimental,paz2009perfect,lau2012proposal}.

In this work, we shall bring this symbiosis further by extending the Exact Holographic Mapping to arbitrary (discrete) wavelets transforms\footnote{Ref. \onlinecite{singh2016holographic} extended it to the family of Daubechies wavelets.}. This more general framework allows for a more physically motivated, basis agnostic interpretation of the bulk geometry, since it explicitly isolates features associated with the choice of wavelet basis. Just as importantly, an EHM based on generic wavelet bases can preserve the functional form of a much larger class of Hamiltonians, in the spirit of conventional RG procedures (The existing EHM based on the Haar wavelet can only preserve linearly dispering Hamiltonians). This will be relevant, amongst various reasons, for the very interesting holographic analysis of topological phases protected by symmetries that also create extra degeneracies in the bandstructure, such as type-II Dirac cones and nodal rings and links\cite{sun2012topological,gao2016classification,lin2016dirac,lee2015negative,yan2016tunable,yan2017nodal,yan2017experimental,bi2017nodal,li20172}. 

This paper is structured as follows. In Section II, we provide a pedagogical introduction to the construction of wavelet bases in a language familiar to physicists, and highlight some properties that play a crucial role in the describing the emergent geometry of the holographic bulk. Following that, we explain in Section III how Hamiltonians are renormalized under the EHM, and how to find the appropriate wavelet basis, if it exist, that keeps a given Hamiltonian invariant. In Section IV, we derive the dependence of the bulk correlators, mutual information and hence bulk geometry on the wavelet basis, focusing on how it arises from the branch cut topology of the boundary propagator. Finally, in Section V we briefly discuss generalizations to other configurations of Fermi points, and also anisotropy in the resultant bulk geometry for multi-dimensional EHM.

\section{Exact Holographic Mapping (EHM) through wavelets}


\subsection{Conceptual overview of the EHM }
The EHM was first introduced in Ref. \onlinecite{qi2013} as a special type of tensor network that implements a lossless RG-type procedure through a hierarchy of local mappings. It was then extended to more than one RG dimension in Ref. \onlinecite{lee2016exact}, where its various mathematical properties were also elaborated.

\begin{figure}
\includegraphics[scale=.27]{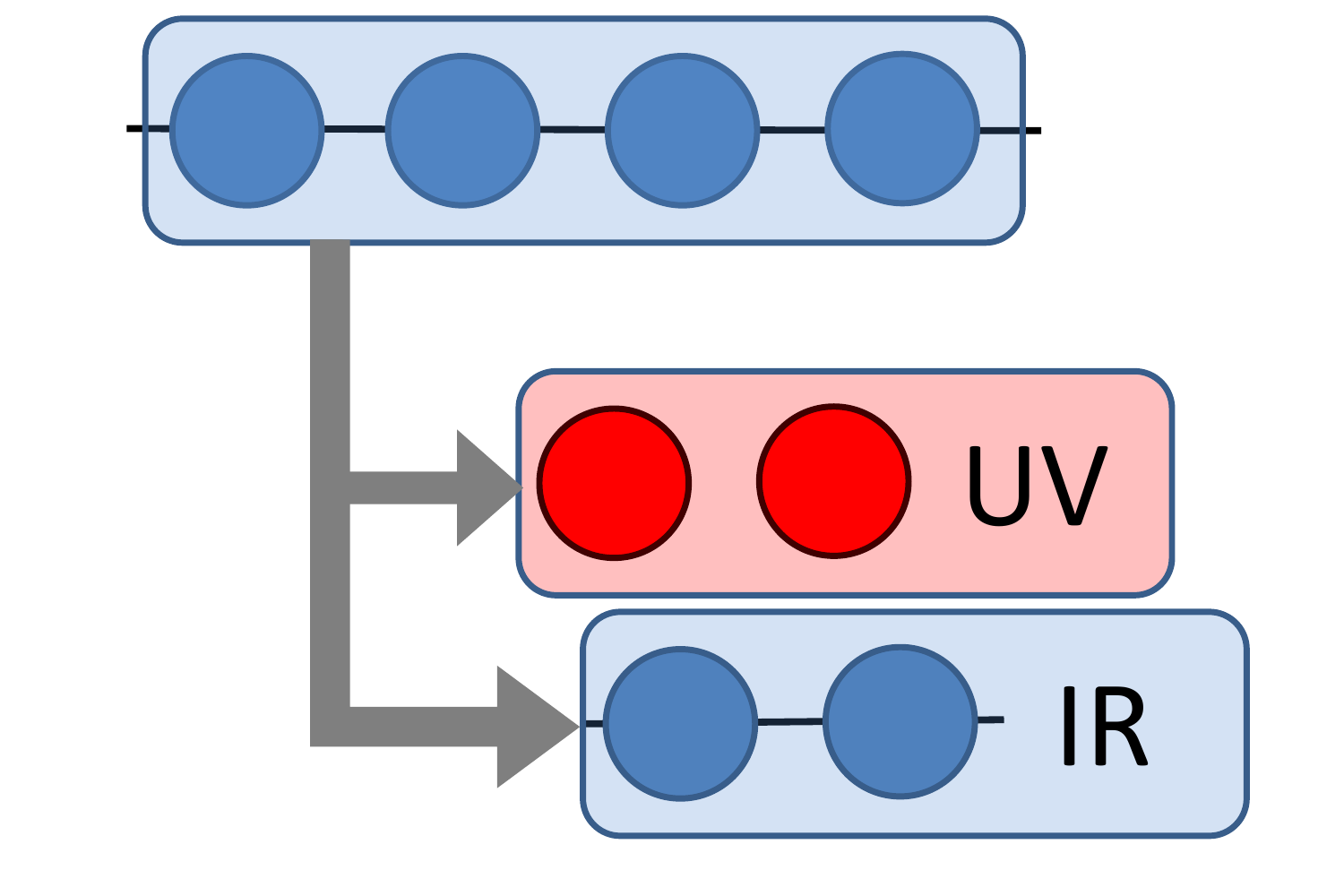}
\includegraphics[scale=.17]{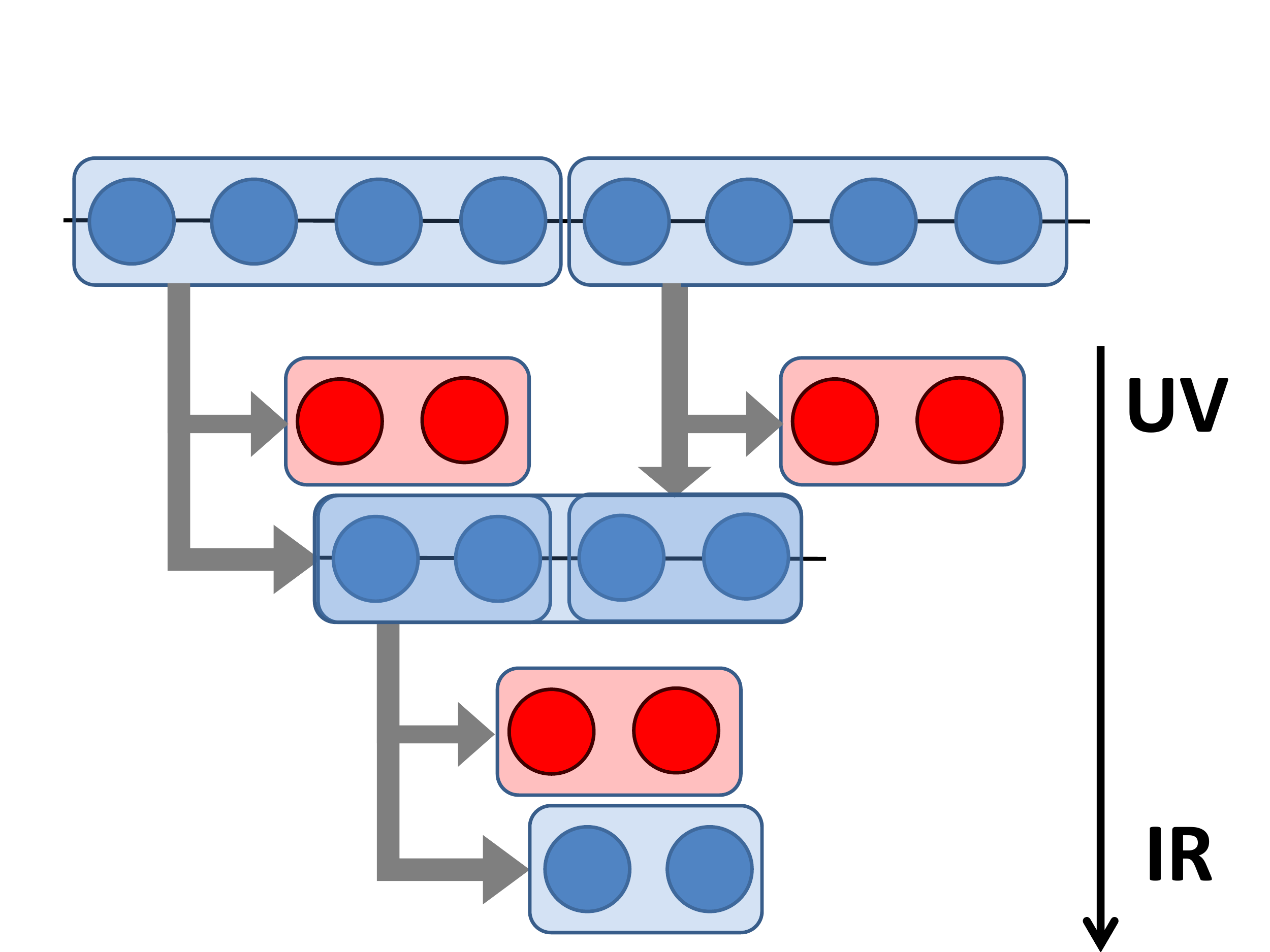}
\caption{Left) Illustration of a single EHM iteration with $l=2$. The degrees of freedom of $2l$ input sites are separated into $l$ UV and $l$ IR sites via a unitary transform. Right) An EHM network with $2$ iteration levels. The IR DOFs from each group of $2l$ sites of the input "boundary" system is fed into the next iteration, until only $l$ sites remain. The collection of the discarded UV (red) DOFs, together with the last remaining IR sites (blue), form the "bulk" system containing the same number of DOFs as the original system.}
\label{fig:wavelet0}
\end{figure}

We start from a given original "boundary" system with $2^Nl$ sites (Fig.~\ref{fig:wavelet0}). At each iteration, the degrees of freedom (qubits) on each $2l$ ($l\geq 1$) adjacent sites are separated into $l$ small-scale (ultraviolet or UV) and $l$ large-scale (infrared or IR) degrees of freedom (DOFs) via a unitary rotation whose form will be elaborated later. 
The $l$ IR DOFs will then be used as the  input for the next iteration, while the UV DOFs will be discarded. This procedure is repeated until we are left with the last set of $l$ sites. 

Since degrees of freedom at larger scales will undergo more iterations before being discarded, the discarded DOFs from all the iterations collectively form an $N+1$-level pyramid-like array arranged hierarchically according to scale.  
We shall define these discarded DOFs as the "bulk" system corresponding to the original "boundary system". Evidently, the bulk system contains the same $2^Nl$ DOFs, but are arranged in levels with $2^{N-1}l$, $2^{N-2}l$, etc. sites according scale.

\subsection{Introduction to wavelets}

The abovementioned EHM procedure is mathematically a discrete\footnote{It is a discrete wavelet transform in the sense that the multiresolution is realized via discrete levels of scale hierarchies. Continuous wavelet transforms lead to an overcomplete though invertible basis description. There also exists an alternative continuous approach to the EHM known as the cMERA\cite{haegeman2013entanglement,mollabashi2013holographic,miyaji2015continuous,gan2016thermal,gan2016emergent,wen2016holographic}. } wavelet transform. Here, we shall provide a pedagogical introduction for its concrete implementation.

\begin{figure*}
\includegraphics[scale=.3]{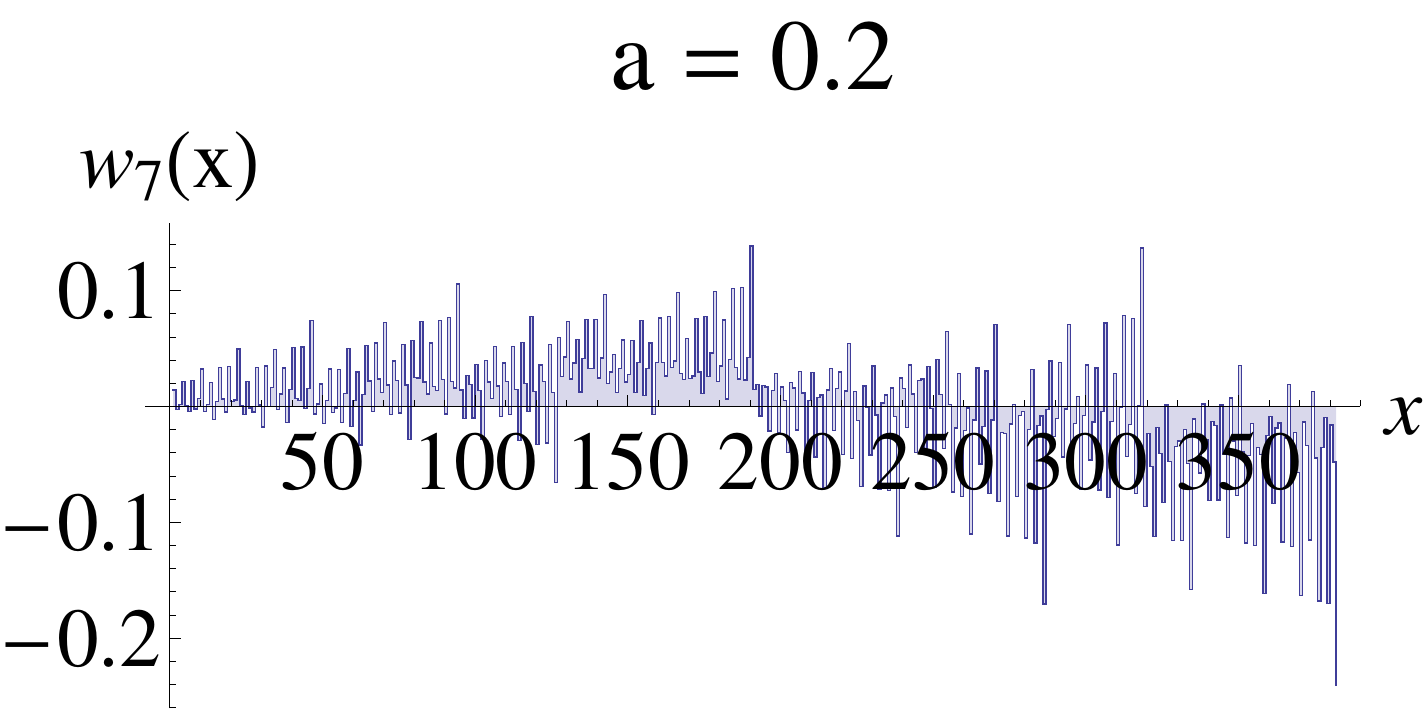}
\includegraphics[scale=.3]{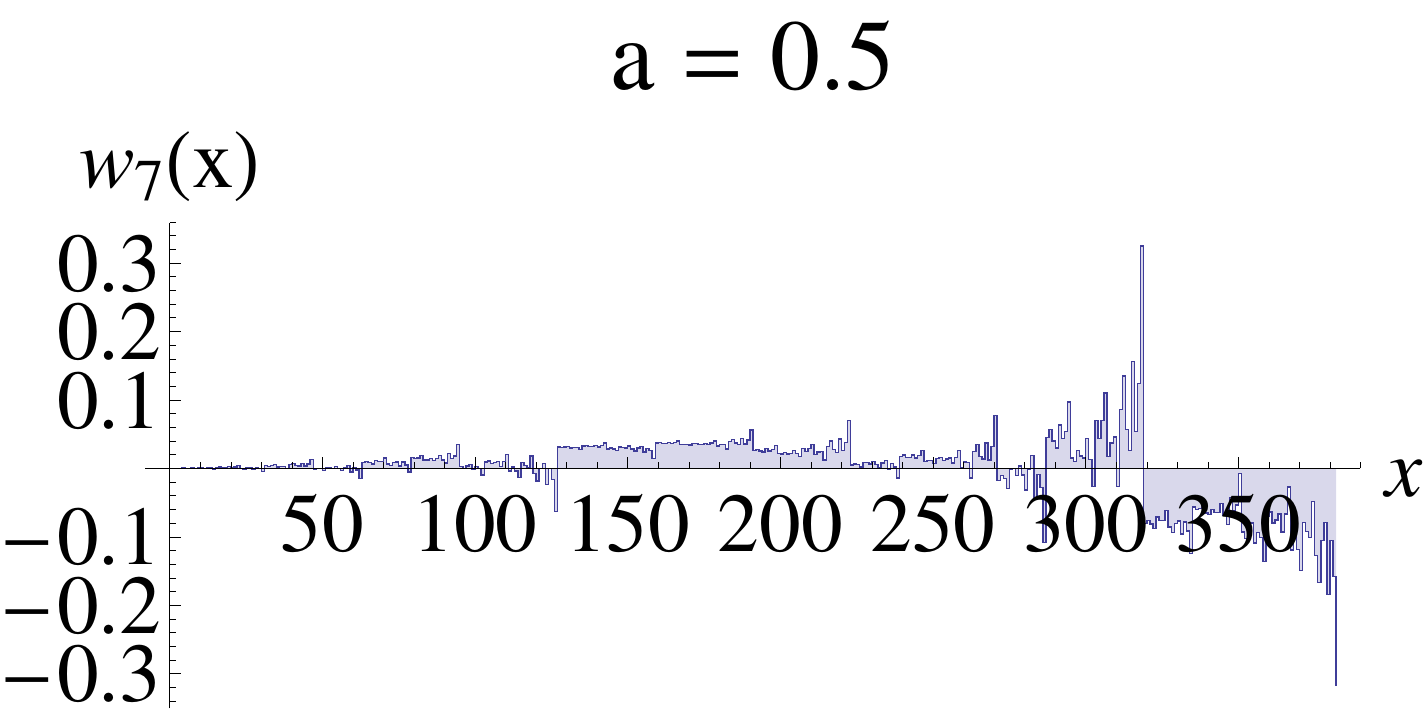}
\includegraphics[scale=.3]{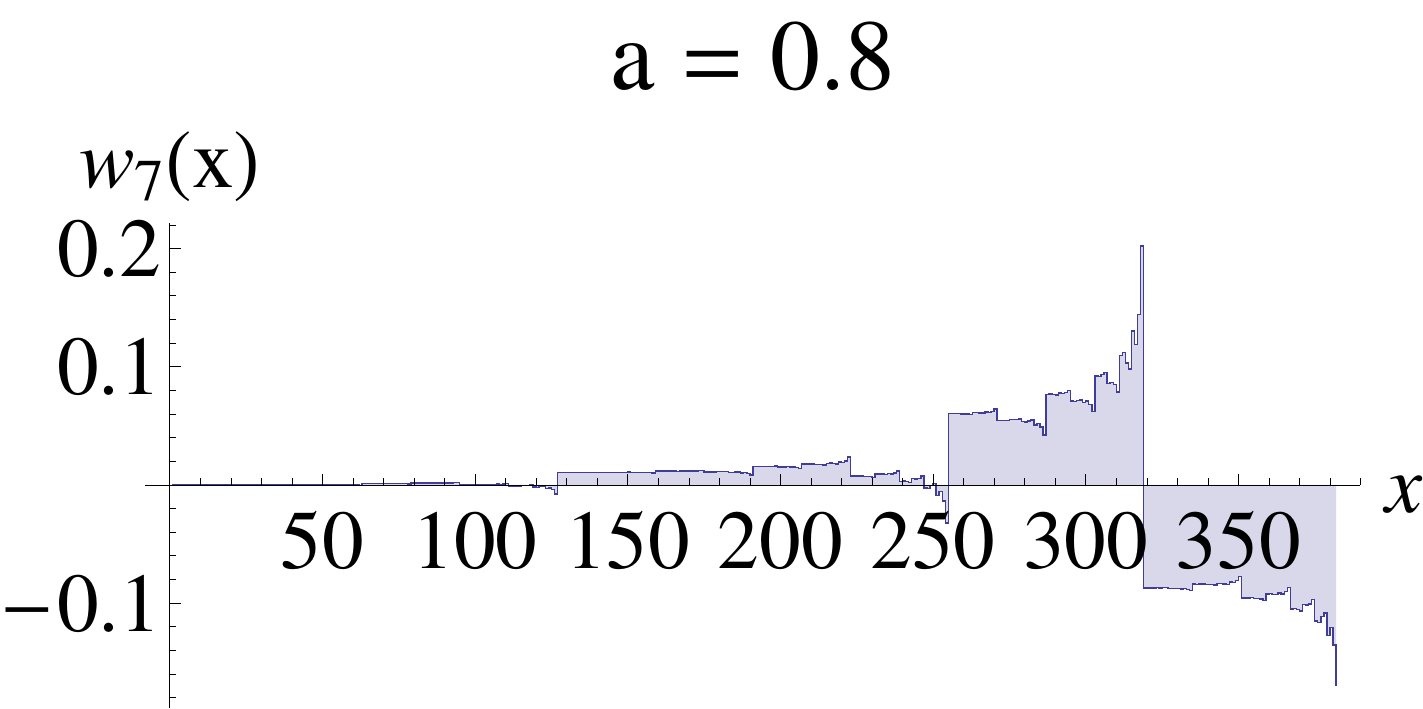}
\includegraphics[scale=.3]{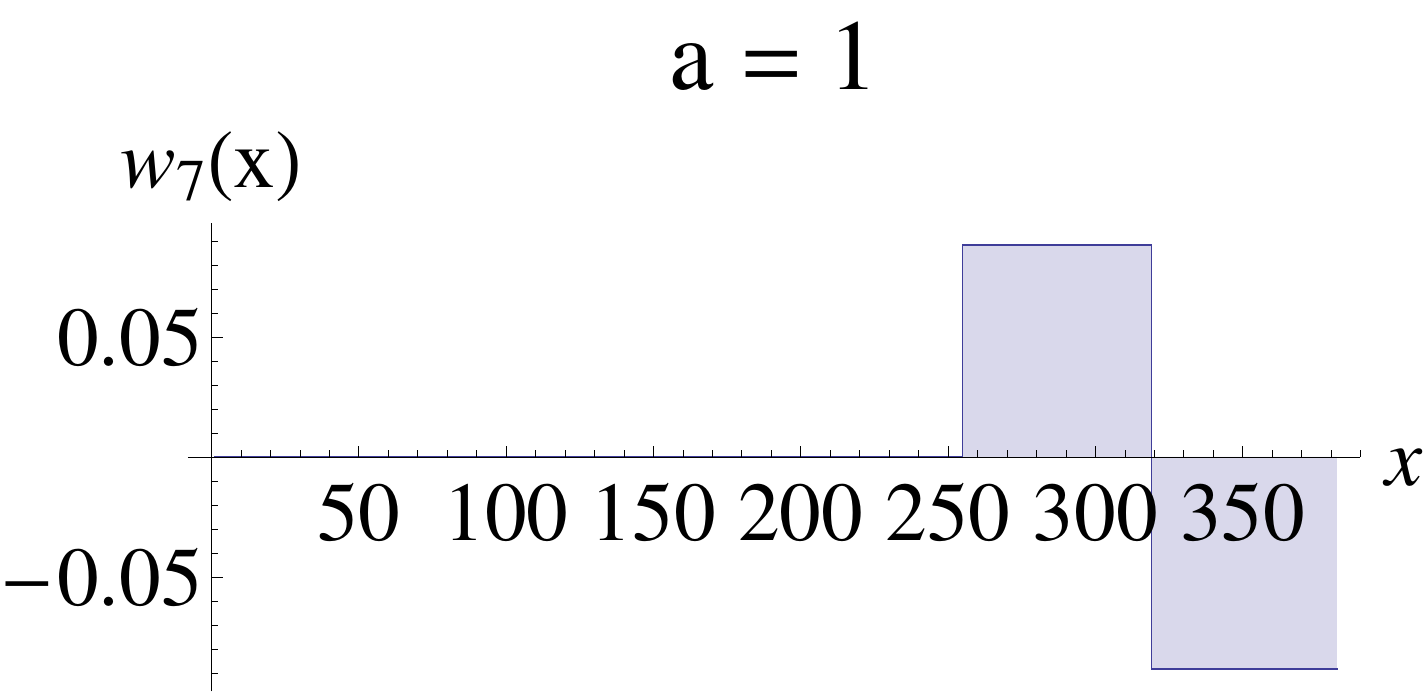}
\includegraphics[scale=.3]{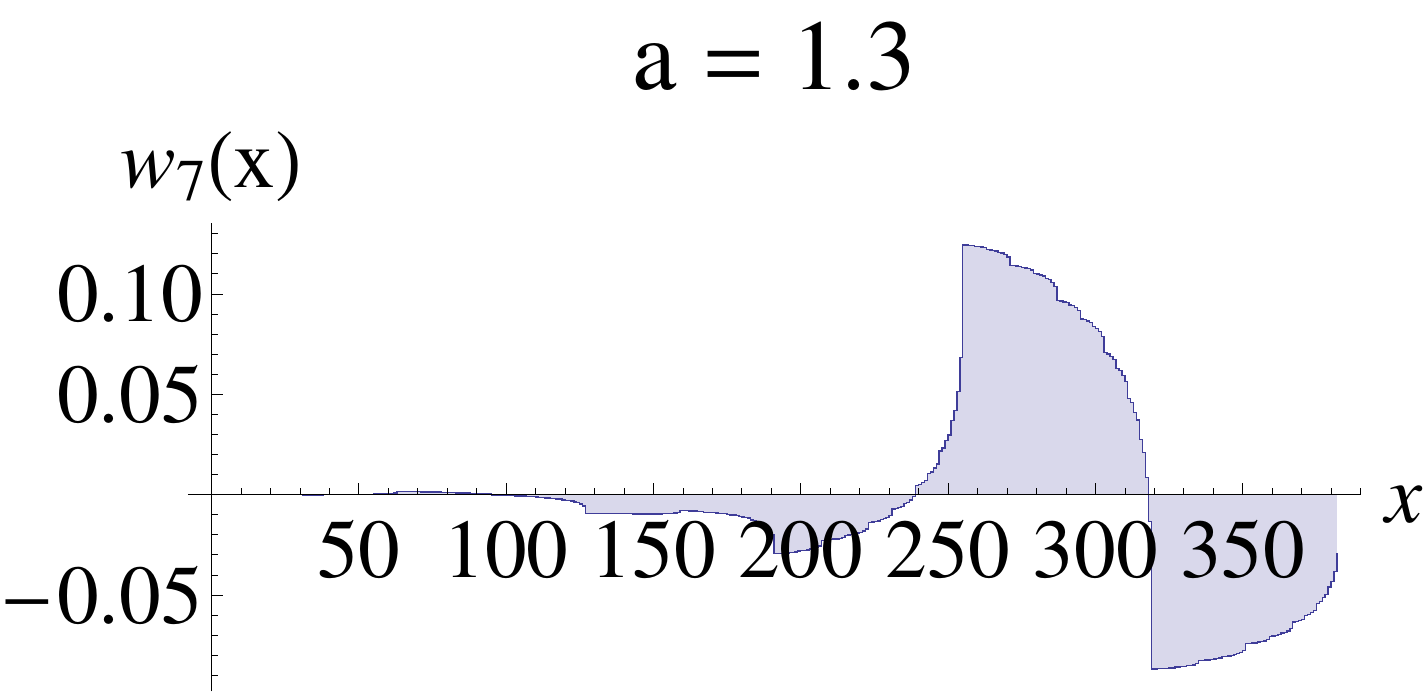}
\includegraphics[scale=.3]{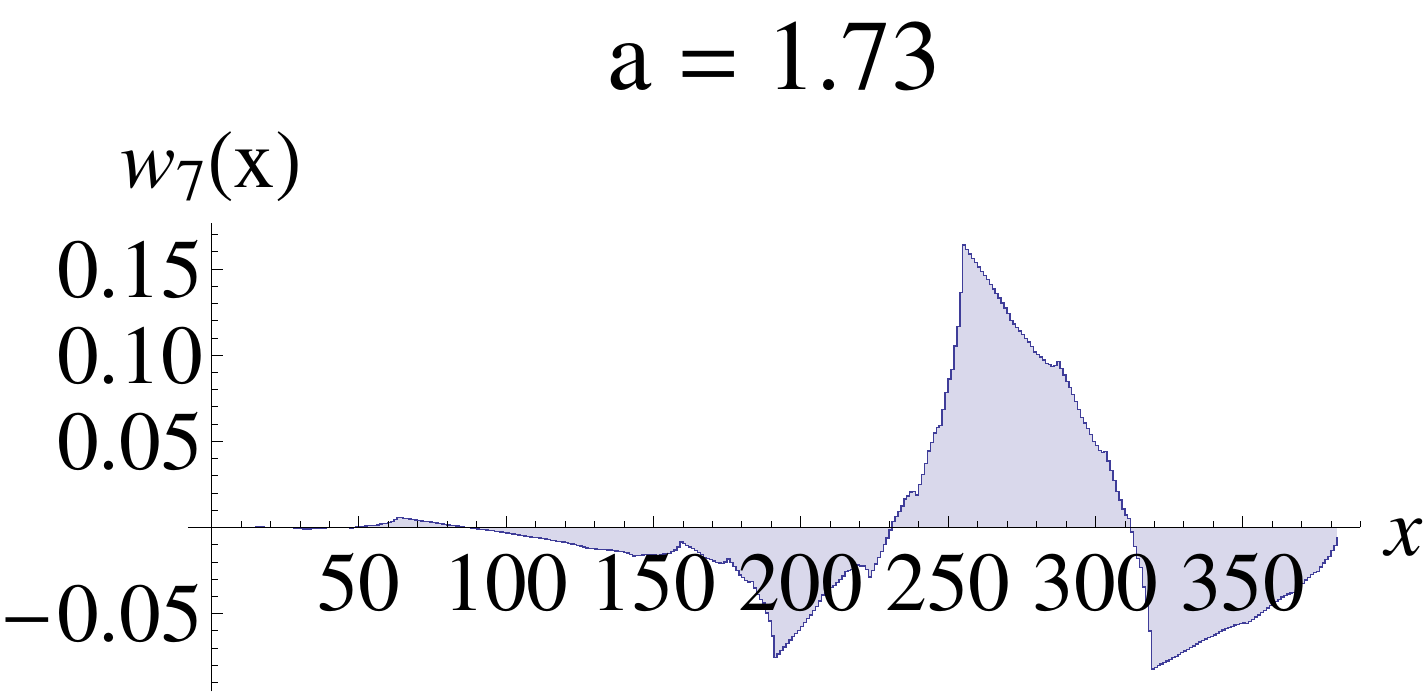}
\includegraphics[scale=.3]{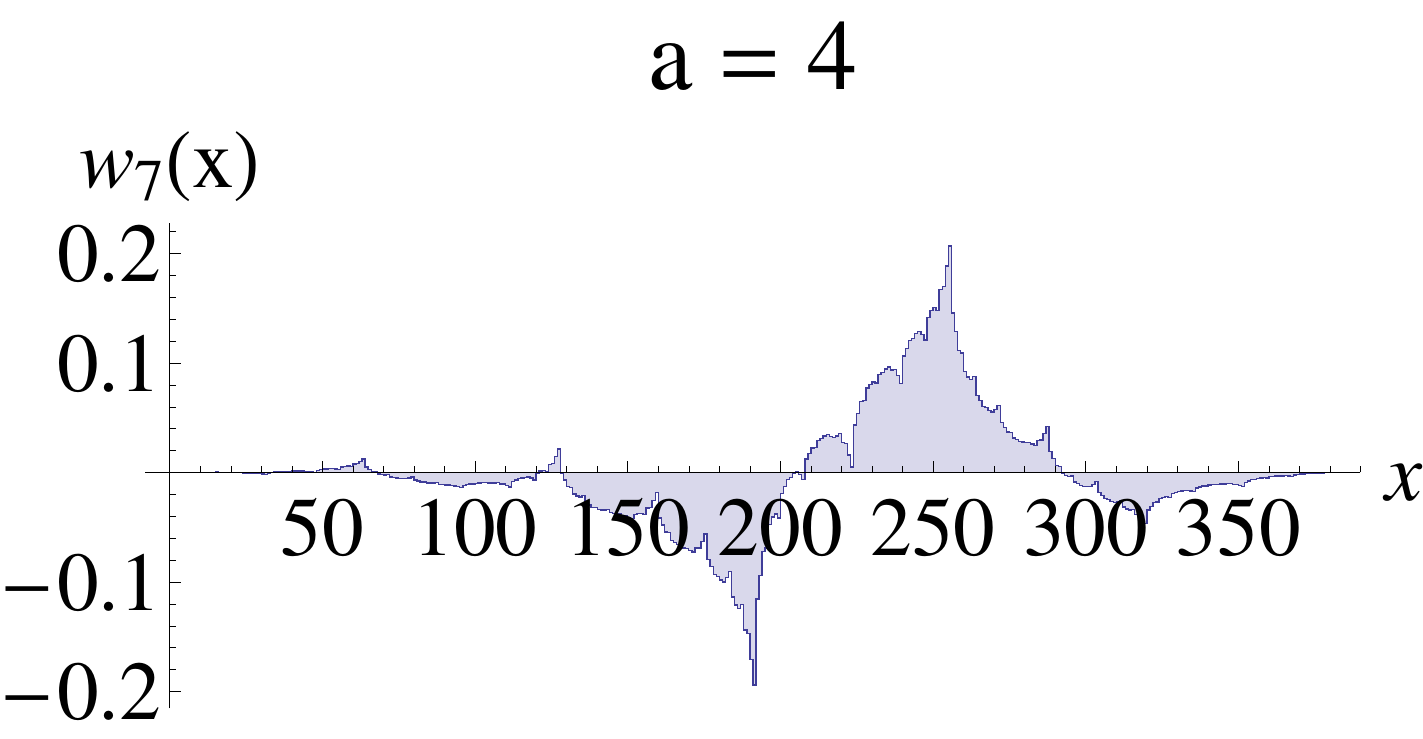}
\includegraphics[scale=.3]{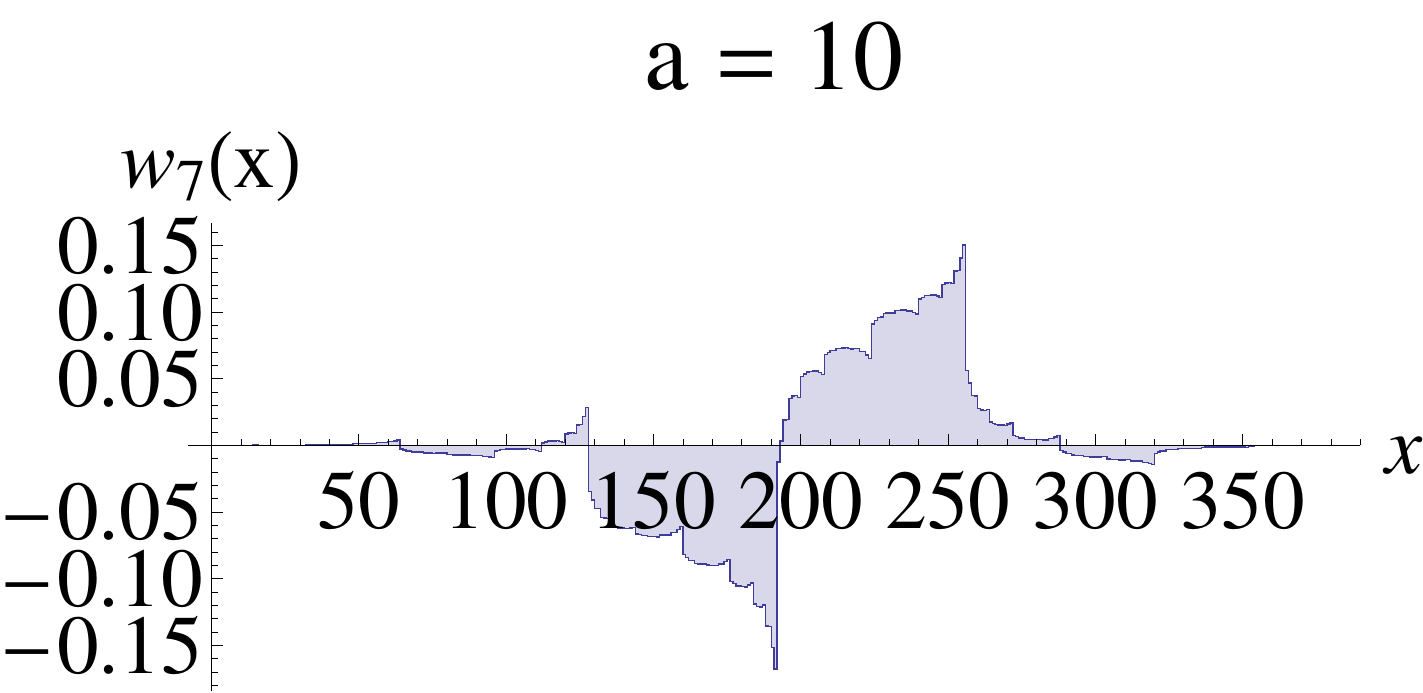}
\caption{Illustration of the real-space wavelets $w_{n,t=0}(x)$ (Eq. \ref{wavelet2}) for various instances with length of unit cell $2l=4$. These wavelets are based on the ansatz $C(z)=\frac{(1-a^*)(1-az)+(1+a)(a^*z^2+z^3)}{\sqrt{2}(1+|a|^2)}$, where $a>0$. It is a more general ansatz than Eq. \ref{Pzalpha}, admitting complex $a$, and can be shown to be consistent with having an odd $P(z)-1$ that is real for $|z|=1$, and which also $P(z)+P(-z)=2$. Physically, $w_{n,t=0}(x)$ represents the "orbital shape" of the UV wavelet basis: In the $a=1$ Haar case for instance, the basis contains two rectangular regions of opposite signs, representing an antisymmetric (short-wavelength) degree of freedom. At other $a$, these basis wavefunctions become either more rounded or jagged. The beauty of wavelet bases is that basis wavefunctions can possess very detailed internal structures, such that only selected features will be "zoomed-in" across wavelet levels. }
\label{fig:wavelet1}
\end{figure*}

A 1-dimensional wavelet system consists of a set of self-similar basis functions defined in exact analogy to the bulk EHM DOFs. It can be described by a scaling function $\phi(x)$ and mother wavelet function $w(x)$ (see Fig. \ref{fig:wavelet1}) pair obeying the recursion relations\cite{daubechies1992}
\begin{equation}
\phi(x)=2\sum_{r=0}^l c(r)\phi(2x-r)
\label{scalingeq}  
\end{equation}
\begin{equation}
w(x)=2\sum_{r=0}^l d(r)\phi(2x-r)  
\label{waveleteq}
\end{equation}
where $d(r)$ and $c(r)$ are the high pass and low pass filter vectors, characterized by spatial fluctuations with shorter and longer length scales respectively. Both $c$ and $d$ are length\footnote{Wavelets filters with finite length $l+1$ are known as finite-impulse response (FIR) filters.} $l+1$ vectors normalized such that $\sum_r|c(r)|^2=|c|^2=|d|^2=1$. In Eq. \ref{scalingeq}, $\phi(x)$ is self-similar in the sense that it is equal to the convolution of $c(r)$ and a rescaled version of itself. The mother wavelet $w(x)$, by contrast, is \emph{not} self-similar, but is the convolution of $d(r)$ and $\phi(2x)$.  In the simplest ($l=1$) case of the Haar wavelet used in Refs. \onlinecite{qi2013,lee2016exact,gulee2016}, we have $c=(1,1)/\sqrt{2}$ and $d=(1,-1)/\sqrt{2}$. 

Through $n$ iterations of Eqs. \ref{scalingeq} and \ref{waveleteq}, one obtains level $n$ wavelets
\begin{equation}
w_{n,t}(x)=w(2^nt-x)  
\label{wavelet2}
\end{equation}
possessing characteristic length scales of $\propto 2^n$. To study the properties of $w_{n,t}(x)$, 
it is useful to define the z-transforms\cite{strang1996}
\begin{align}
C(z)&=\sum_{r=0}^l c(r)z^r\\
D(z)&=\sum_{r=0}^l d(r)z^r,
\end{align}
such that $C(z),D(z)$ with $z=e^{ik}$ are the Fourier transforms of the low pass and high pass filters respectively. (For the whole of this paper, we shall use the same symbol for a function whether its argument is given by $k$ or $z=e^{ik}$) The RG properties of the EHM are most succinctly described by the spectral properties of these filters. For future reference, we shall denote by $C^*$ and $D^
*$ the polynomial $C,D$ with coefficients (but not the argument $z$) conjugated.

The possible choices for filters polynomials $C(z)$ and $D(z)$ are constrained by biorthogonality, that is, by the requirement that $\phi(x)$ and $w(x)$ should be orthogonal to their translates and among themselves. 

For instance, the constraint $(\phi(x),w(x+x_0))=0$ where $x_0\in \mathbb{Z}$ stipulates that the low pass and high pass filters project onto orthogonal subspaces. This requires that
\begin{eqnarray}
0&=& \sum_x\phi^*(x)w(x+x_0)\notag\\
&\propto &\sum_x\sum_{r,r'}c^*(r)d(r')\phi^*(2x-r)w(2x+2x_0-r')\notag\\
&\propto &\sum_{r,r'}c^*(r)d(r')\delta_{r,r'-2x_0}\notag\\
&=& \sum_r c^*(r)d(r+2x_0)
\label{ortho1}
\end{eqnarray} 
which implies that
\begin{equation}
0=\sum_k e^{2ix_0k}C^*(e^{-ik})D(e^{ik})=\frac{1}{2\pi i}\oint \frac{C^*(z^{-1})D(z)dz}{z^{1-2x_0}}.\label{CD}\end{equation}
By the residue theorem, $C^*(z^{-1})D(z)$ must hence have no term with even power, including the constant term. This can be guaranteed by the alternating-flip construction $d(r)=(-1)^rc(l-r)$, i.e.
\begin{equation}
C(z)=z^l D\left(-\frac1{z}\right)
\end{equation}
where $l$ is the degree of the polynomials $C(z)$ and $D(z)$. 

Also, $\phi(x)$ and a translated copy of itself $\phi(x+x_0)$, $x_0\in \mathbb{Z}/\{0\}$ should be orthogonal in order to form a local basis. This requires that
\begin{eqnarray}
\delta_{x_0,0}&=& \sum_x\phi^*(x)\phi(x+x_0)\notag\\
&\propto &\sum_x\sum_{r,r'}c^*(r)c(r')\phi^*(2x-r)\phi(2x+2x_0-r')\notag\\
&\propto &\sum_{r,r'}c^*(r)c(r')\delta_{r,r'-2x_0}\notag\\
&=& \sum_r c^*(r)c(r+2x_0),
\label{ortho2}
\end{eqnarray} 
so $C^*(z^{-1})C(z)$ has a constant term of $1$, but no non-constant term with even power. An analogous constraint holds for $D(z)$. Since the latter is a high-pass filter, it should satisfy the additional constraint that it has zero weight in the long wavelength limit $k=0$ (or $z=1$). As such, $D(1)=\sum_r d(r) e^{i 0\cdot r}=\sum_r d(r)=0$ (But see Sect. \ref{sec:zoomin} for a reason to break this constraint).

All in all, the wavelet basis is \emph{completely} determined by the autocorrelation Laurent polynomial
\begin{eqnarray}
P(z)&=&C^*(z^{-1})C(z)=D(-z^{-1})D^*(-z)\notag\\
&=&1+\sum_{j=1} \left[ p_{2j-1}z^{2j-1}+\frac{p^*_{2j-1}}{z^{2j-1}}\right]
\label{Pz}
\end{eqnarray}
whose coefficients $p_{2j-1}$ take values such that $P(z)\geq 0$ for all $|z|=1$ on the unit circle, and normalized such that $P(1)=2$. The absence of nontrivial even powers of $z$ also implies that $P(z)+P(-z)=2$. $C(z)$ and $D(z)$, which are related by Eq. \ref{CD}, can be obtained via a factorization\footnote{This factorization can in general be accomplished by numerical methods like the Cepstral method or Wiener-Hopf factorization. See chapter 5.4 of Ref. \onlinecite{strang1996} for an introduction.} of $P(z)$.  

We are now ready to derive specific allowed forms for the wavelet functions. Since $\phi(x)$ is a convolution of $c(r)$ and $\phi(2x)$ (Eq. \ref{scalingeq}), its z-transform obeys
\begin{eqnarray}
\Phi(z)&=&\Phi(\sqrt{z})C(\sqrt{z})\notag\\
&=&\Phi(z^{1/4})C(z^{1/4})C(z^{1/2})\notag\\
&=&\;...\;=\prod_{b=1}^\infty C\left(z^{2^{-b}}\right)
\label{scaling2}
\end{eqnarray}
This is the explicit expression for the (z-transform of the) scaling function $\Phi$ in terms of its recursive definition. Of course, the infinite product should terminate finitely when we are in a discrete system. For that, we can obtain from Eqs. \ref{waveleteq} and \ref{scaling2} wavelet spectral functions $W_n(z)$ corresponding to the wavelets $w_n(x)$ at scale level $n$:  
\begin{eqnarray}
W_n(z)&=& \frac1{\sqrt{2\pi}}W\left(z^{2^n}\right) \notag\\
&=& \frac1{\sqrt{2\pi}}D\left(\sqrt{z^{2^n}}\right)\Phi\left(\sqrt{z^{2^n}}\right)\notag\\
&=& \frac1{\sqrt{2\pi}}D\left(z^{2^{n-1}}\right)\prod_{b=0}^{n-2}C\left(z^{2^b}\right)
\label{wavelet2}
\end{eqnarray} 
where the additional normalization factor of $\frac1{\sqrt{2\pi}}$is introduced for future notational consistency. Hence the construction of a (1-dimensional) wavelet basis involves these three basic steps:
\begin{enumerate}
\item Choosing a polynomial $P(z)=P(e^{ik})$ with desired spectral properties (Eq. \ref{Pz}).
\item Factorization of $P(z)$ into $C(z)$ and $D(z)$.
\item Construction of wavelet spectral functions $W_n(z)$ via Eq. \ref{wavelet2}.
\end{enumerate}
As a simplest illustration, the Haar wavelet is characterized by $P(z)=1+\frac{z+z^{-1}}{2}$, which factorizes to $C(z)=\frac{1+z}{\sqrt{2}}$, $D(z)=\frac{1-z}{\sqrt{2}}$. From Eq. \ref{wavelet2}, the Haar wavelet spectral functions are thus given by $W_n(z)=\sqrt{2^-n}(1-z^{2^{b-1}})\prod_{b=0}^{n-2}\left(1+z^{2^b}\right)=2^{-n/2}\frac{\left(1-z^{2^{n-1}}\right)^2}{1-z}$. This is illustrated by the $\alpha=1$ case shown in Fig. \ref{fig:wavelet2}.


Finally, one can compute the spectral weight $|W_n(z)|^2$ of each wavelet level directly through the autocorrelation function: 
\begin{eqnarray}
|W_n(z)|^2&=&W^*_n(z^{-1})W_n(z)\notag\\
&=& \frac1{2\pi}P\left(-z^{-2^{n-1}}\right)\prod_{b=0}^{n-2}P\left(z^{2^b}\right)
\end{eqnarray}

\subsection{Implementation of wavelets in the EHM}
The Exact Holographic Mapping is most easily understood in terms of its wavelets in momentum space. Writing the second quantized operators of the original (boundary) system as $a^\dagger_k=\frac1{\sqrt{2^Nl}}\sum_x e^{ikx}a^\dagger_x$, the EHM is just a unitary transform to the basis of (bulk) states created by 
\begin{equation}
b^\dagger_{nx}=\sum_k W_n^*(e^{-ik})e^{-i2^nkx}a^\dagger_k
\label{EHMbasis}
\end{equation}
where $n\geq 1$ indexes the level and $x=1,2,...,2^{N-n}l$ denotes the position within level $n$. Hence the original $2^Nl$ DOFs $a^\dagger_x|0\rangle$ are re-distributed into a pyramid with $2^{N-n}l$ sites (DOFs) $b^\dagger_{nx}|0\rangle$ at level $n$ (Fig. \ref{fig:wavelet0}). Note that $k$ refers to the momentum defined \emph{within each} level: On the $n^{th}$ level with $2^{N-n}l$ sites, $k=\frac{2\pi j}{2^{N-n}l}$ where $j\in \mathbb{Z}$. That Eq. \ref{EHMbasis} represents a unitary transformation of the \emph{Hilbert space} can be seen from the biorthogonality of $W_n(z)z^{2^nx}$, which is proven in Appendix \ref{sec:biortho}.

\subsection{Wavelet properties relevant to holography}
\label{sec:holowav}
We have seen that a wavelet basis naturally provides a way to decompose information into a hierarchy of basis vectors at various scales. Furthermore, these wavelet bases are local and thus suitable candidates for describing physical degrees of freedom in real space. This should be contrasted with Fourier transforming into the momentum space basis, where each momentum mode is periodic and not compactly supported.

Below, we highlight a few properties of the wavelet basis that play a key role in the EHM. Of most significance is the smoothness of the IR filter $C(z)$ in the long wavelength limit $z=1$ (or $k=0$). This smoothness is characterized by an integer $\kappa$, which is the order of the first nonzero derivative (number of vanishing moments) of $C(z)$ at $z=1$, i.e. $C^{(\kappa)}(1)\neq 0$ but $C^{(\kappa')}(1)=0$ for $\kappa'<\kappa$. Equivalently, $P(z)$ has $2\kappa-1$ vanishing moments.
\begin{figure}[H]
\includegraphics[scale=.43]{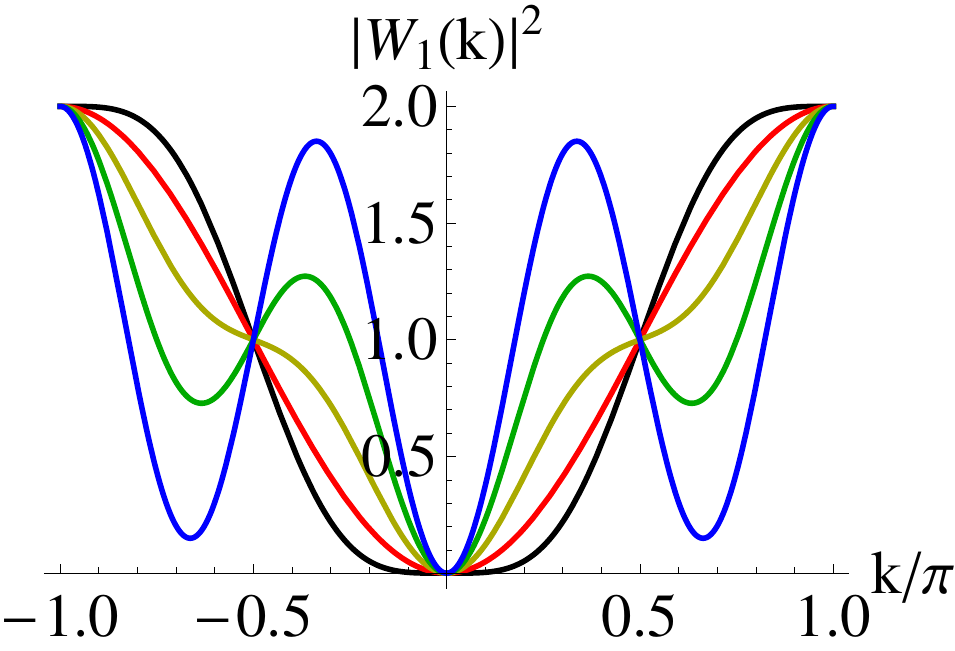}
\includegraphics[scale=.43]{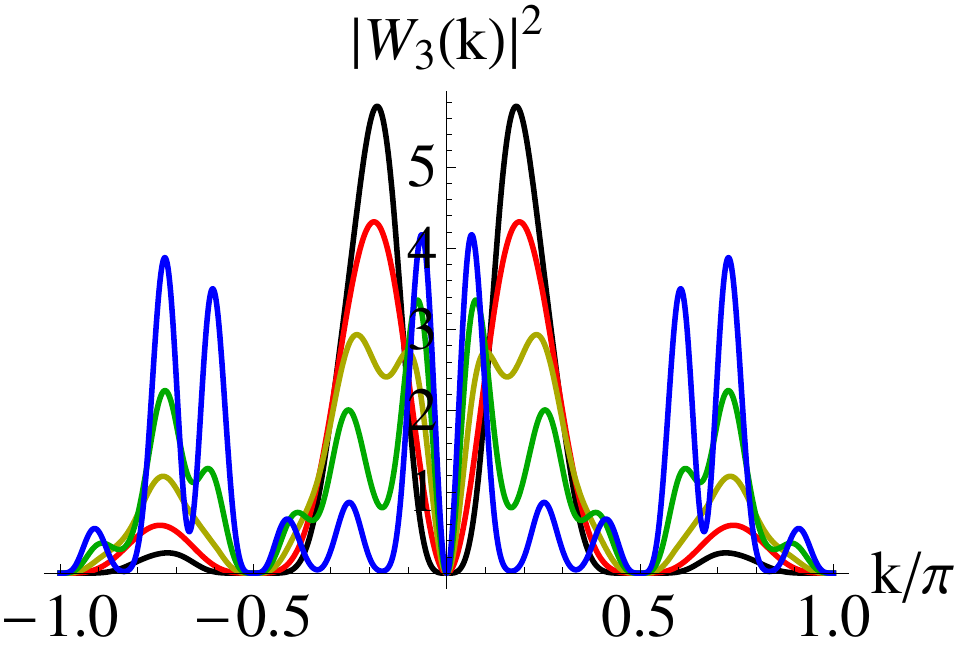}
\includegraphics[scale=.43]{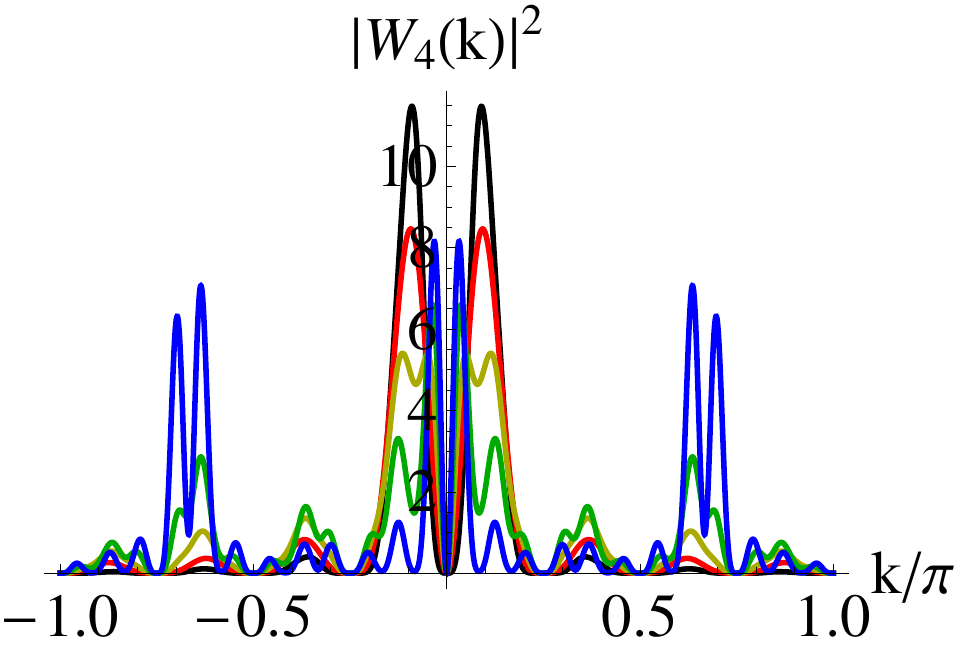}
\includegraphics[scale=.43]{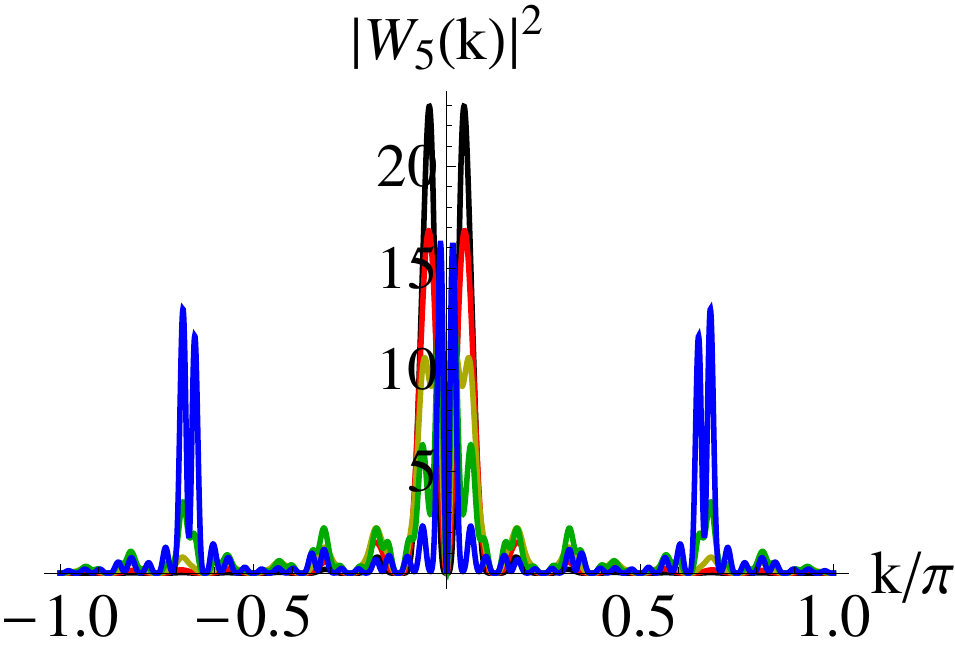}
\caption{The spectral weights $|W_n(e^{ik})|^2$ for levels $n=1,3,4,5$ (Top Left, Top Right, Bottom Left, Bottom Right). Plotted in each figure are the spectral weights corresponding to $l=2$ wavelets described by Eq. \ref{Pzalpha} for $\alpha=1.25,1,0.6,0,-0.8$ (Black, Red,Yellow,Green and Blue). The $\alpha=1$ case (Red) corresponds to the simplest Haar wavelet with no $z^3$ dependence. The next lowest Daubechies wavelet is given by $\alpha=1.25$ (Black), and has the special property that of having $2\kappa-1=3$ vanishing derivative orders (moments) of $P(w)$ at $k=0$. Consequently, with contributions away from the peaks most strongly suppressed, it has the strongest peaks among all the other $\alpha$. As $\alpha$ decreases, the IR DOFs become less effectively suppressed, leading to higher secondary peaks.}
\label{fig:wavelet2}
\end{figure}
The significance of $\kappa$ is illustrated in Fig. \ref{fig:wavelet2}, where the spectral weights $|W_n(e^{ik})|^2$ for levels $n=1,3,4$ and $5$ are plotted for $P(z)$ of the form
\begin{equation}
P(z)=1+\frac{1+\alpha}{4}\left(z+z^{-1}\right) +\frac{1-\alpha}{4}\left(z^3+z^{-3}\right)
\label{Pzalpha}
\end{equation}
One readily checks that $P(1)=2$, $P(z)+P(-z)=2$ and $P(z)\geq 0$ for $-1<\alpha <\frac{5}{4}$. For the special case of $\alpha=\frac{5}{4}$, $P(z)$ factorizes to $\left(1+\frac{z+\frac1{z}}{2}\right)^2\left(1-\frac{z+z^{-1}}{4}\right)=(1+\cos k)^2\left(1-\frac{\cos k}{2}\right)$, i.e. $P(1)=P'(1)=P^{(2)}(1)=P^{(3)}(1)=0$, implying that $\kappa =2 $. This case is represented by the black curve in Fig. \ref{fig:wavelet2}, which possesses a spectral weight that is strongly suppressed at $k=0$ even for the first level $n=1$. This strong suppression is further magnified in subsequent levels, with the corresponding $D\left(z^{2^{n-1}}\right)$ factor giving rise to the sharpest IR peaks compared to the other cases with fewer vanishing moments, i.e. $\kappa =1$.

\begin{figure}[H]
\includegraphics[scale=.33]{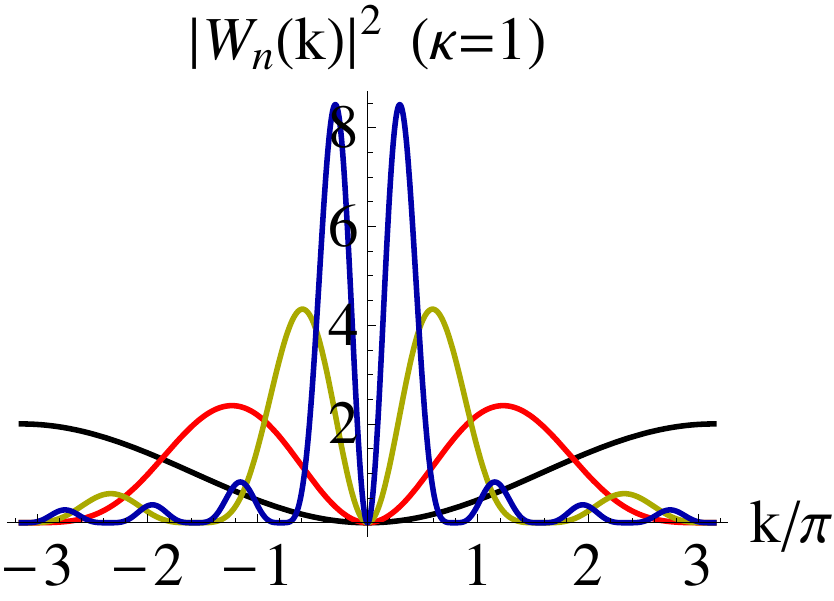}
\includegraphics[scale=.33]{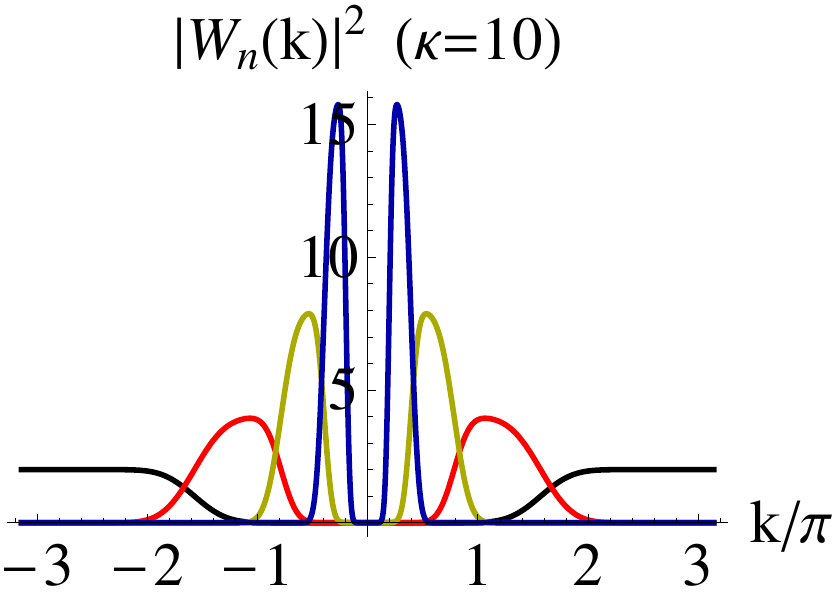}
\includegraphics[scale=.33]{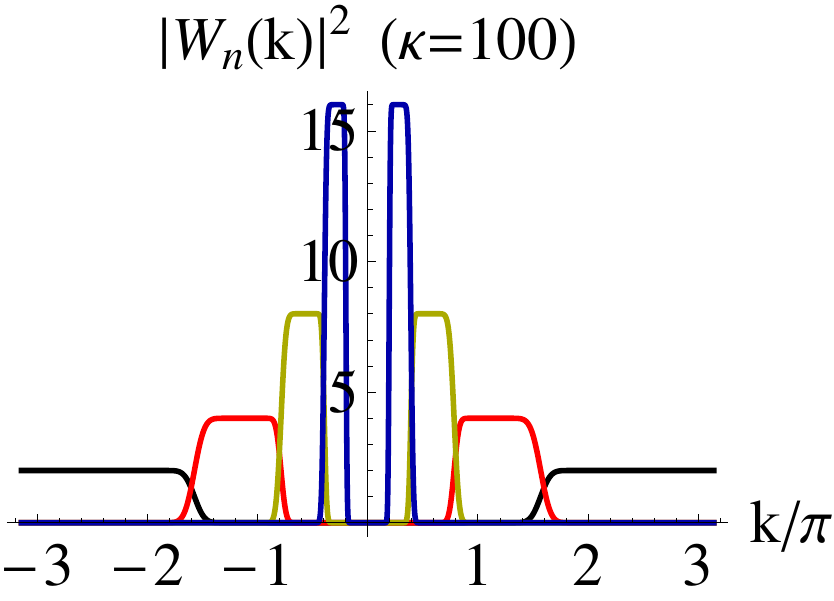}
\caption{Spectral weight of levels $n=1,2,3,4$ of the Daubechies family of wavelets for $\kappa=1,10$ and $100$. We see extremely smoothness at $k=0$ in the large $\kappa$ limit, since the wavelet filter has a zero with $\kappa-1$ vanishing moments. In real space, this extreme smoothness with momentum corresponds to extremely narrow peaks.}
\label{fig:wavelet3}
\end{figure}

In general, wavelet mappings with higher $\kappa$ are more effective at suppressing DOFs away from the limiting IR point, and thus have more pronounced spectral peaks at $k=\pm\frac{2\pi}{2^n}$ at the $n^{th}$ level. In essence, wavelet mappings represent a trade-off between locality and sharpness of scale resolution: A sharp momentum cutoff requires non-local (power-law decaying) real space components, while the most local mapping (the Haar wavelet) lead to rounded spectral peaks. With a given length $2l$ for the mother wavelet, the maximal $\kappa$ and hence best possible spectral resolution is realized by the Daubechies' wavelet family with $P(z)$ (Fig. \ref{fig:wavelet3}) given by
\begin{eqnarray}
&&P_{Daub}(z)\notag\\
&=&2\left(1+\frac{z+z^{-1}}{2}\right)^l\sum_{j=0}^{l-1}\binom {l+j-1}{j}\left(1-\frac{z+z^{-1}}{2}\right)^j/2^{l+j}\notag\\
\label{Daub}
\end{eqnarray}
which reproduces the abovementioned $\alpha=\frac{5}{4}$ wavelet when $l=2$, and the Haar wavelet when $l=1$. That $P_{Daub}(z)$ has $\kappa=l$, the maximum possible value for a given $l$, can be seen\cite{daubechies1992,strang1996} by expressing it in terms of $y=\frac1{2}-\frac{z+z^{-1}}{4}$, which yields $P_{Daub}'(y)\propto y^{l-1}(1-y)^{l-1}$. 

\section{Renormalization of Hamiltonians under the EHM}

Regarded as a lossless renormalization group (RG) procedure, the Exact Holographic Mapping should ideally preserve the form of the Hamiltonian under renormalization. Below we shall discuss when this is possible, and how can the renormalization scale parameter be determined. This will greatly generalizes the scope of previous literature\cite{qi2013, lee2016exact}, where the special choice of the Haar wavelet basis preserves the form of Dirac-type Hamiltonians $\sin k \sigma_1+(m+1-\cos k)\sigma_2$ only. 


Let $h^n$ be the input of the $n^{th}$ EHM iteration of the original Hamiltonian $h$. From Eq. \ref{scaling2}, $h^{n+1}$ is related to $h^n$ via a multiplication with the wavelet spectral weight $|C|^2$. Writing $h(w)$ as $h(k/2)$ (with a slight misuse of notation), such that $w=e^{ik/2}$, we have
\begin{eqnarray} &&2h^{n+1}(w)\notag\\
&=& h^n(w)C(w)C^*\left(w^{-1}\right) + h^n(-w) C(-w)C^*\left(-w^{-1}\right) \notag\\
&=& \sum_{\pm} h^n(\pm w)P(\pm w) 
\label{RG0}
\end{eqnarray}
The two copies of momenta $w=e^{ik/2}$ and $-w=e^{i(k+\pi)/2}$ in the summation arise due to a folding of the Brillouin zone, since level $n$ has twice as many sites as level $n+1$. Hence, we have $h^{n+1}(w)$ given by the average of $h(\pm w)$ weighted by the wavelet autocorrelation function from both $\pm w$. 

To find conditions on the wavelet that leaves the Hamiltonian invariant, we set $h^n$ and $h^{n+1}$ in Eq. \ref{RG0} to have the same functional form $h$:
\begin{eqnarray}
\lambda h(w^2) 
&=&\frac1{2}\sum_{\pm} h(\pm w)P(\pm w)\notag\\
&=& h_{even}(w)+ [P(w)-1]h_{odd}(w)\notag\\
\label{RG2}
\end{eqnarray}
where $\lambda$ is the (constant) scale factor for each RG step, and $h_{even/odd}(w)=\frac1{2}\left(h(w)\pm h(-w)\right)$. In other words, given a Hamiltonian $h(w)=h_{even}(k)+h_{odd}(k)$, the wavelet that fixes it must have the autocorrelation function 
\begin{eqnarray}
P(w)=C^*(z^{-1})C(z)=1+\frac{\lambda h(w^2)-h_{even}(w)}{h_{odd}(w)}
\label{RG3}
\end{eqnarray}
Here are a few caveats about Eq. \ref{RG3}:
\begin{enumerate}
\item The RG scale factor can only take nontrivial values of $\lambda\neq 1$ if the Hamiltonian is gapless (critical) in the long-wavelength limit $k=0$ ($w=1$). This follows immediately by setting $w=1$ and noting that $P(1)=2$. 
\item Given $h(w)$ of degree $d$, the degree of $P(w)$ (or $C(w)$) is fixed by comparing the leading powers of Eq. \ref{RG2} to be $l=2\lfloor \frac{d}{2}\rfloor+1$ 
\item There may not exist a wavelet that leaves the form of a given $h(w)$ invariant. Existence of the former is contingent on the RHS of Eq. \ref{RG3} being factorizable into an odd Laurent polynomial $P(w)$ with odd powers $-l$ to $l$, such that it is real for $|w|=1$, and that $P(w)+P(-w)=2$ (i.e. of the form Eq. \ref{Pz}). Further discussion is given in Appendix \ref{sec:matrix}; refer to the next subsection for specific examples of invariant Hamiltonians and their associated wavelets.

\end{enumerate}
In the critical case $h(1)=0$, $\lambda$ is determined by the constraint $P(1)=2$. Eq. \ref{RG3} gives 
\begin{eqnarray}
\lambda&=&\lim_{w\rightarrow 1}\frac{h_{even}(w)+ [P(w)-1]h_{odd}(w)}{h(w^2)}\notag\\
&=&\lim_{w\rightarrow 1}\frac{h'_{even}(w)+ [P(w)-1]h'_{odd}(w)+P'(w)h_{odd}(w)}{2wh'(w^2)} \notag\\
&=&\frac1{2}+\lim_{w\rightarrow 1}\frac{[P(w)-2]h'_{odd}(w)+P'(w)h_{odd}(w)}{2wh'(w^2)} \notag\\
\label{RG4}
\end{eqnarray}
If $h'(1)\neq 0$, the limit on the last line is easily taken and $\lambda=\frac{1}{2}$, unless $P'(1)$ and $h_{odd}(1)$ are both nonzero. This can occur only if $P(w)$ do not have real coefficients, and $h(w)$ is neither odd nor even. Letting $\gamma$ be the order of the first nonzero derivative of the Hamiltonian $h(w)$ at $w=1$, we have
\begin{equation}
\lambda|_{\gamma=1}=\frac1{2}\left(1+\frac{P'(1)h_{odd}(1)}{h'(1)}\right)
\label{lambda1}
\end{equation}
Frequently, the Hamiltonian is not linearly dispersive at $w=1$, and to evaluate $\lambda$ we will need to invoke L'H\^{o}pital's rule a total of $\gamma$ number of times. For $\gamma=2$, we get 
\begin{equation}
\lambda|_{\gamma=2}=\frac1{4}\left(1+\frac{P''(1)h_{odd}(1)+2P'(1)h'_{odd}(1)}{h''(1)}\right)
\label{lambda2}
\end{equation}
and, in general,
\begin{eqnarray}
\lambda|_{\gamma}&=&\frac1{2^\gamma}\left(1+\frac{\sum_{j=1}^{\gamma}P^{(j)}(1)h^{(3-j)}_{odd}(1)}{h^{(\gamma)}(1)}\right)\notag\\
&=&\frac1{2^\gamma}\left(1-\frac{\sum_{j=1}^{\gamma}P^{(j)}(1)h^{(3-j)}_{even}(1)}{h^{(\gamma)}(1)}\right)
\label{lambda3}
\end{eqnarray}
As such, an EHM iteration rescales the Hamiltonian by a factor of $\frac1{2}$ for each vanishing order of $h(w=1)$ if it is either fully even or odd. Otherwise, $\lambda$ will be more complicated, depending on the derivatives of the resultant wavelet autocorrelation $P(w)$. 


\subsection{Renormalization examples}
\subsubsection{Simplest case: Haar wavelet}

With the Haar wavelet basis, $C(w)=\frac{1+w}{\sqrt{2}}$ and $P(w)=C(w)C^*(w^{-1})=1+\frac{w+w^{-1}}{2}$. It is easy to verify that the two linearly independent solutions to Eq. \ref{RG2} are $h(w)= \frac{w-w^{-1}}{2i}\Rightarrow h(k)=\sin k$ and 
$h(w)=\frac{2-w-w^{-1}}{4}\Rightarrow h(k)=\frac{1-\cos k}{2} $, both with the RG rescaling $\lambda =\frac{1}{2}$, consistent with Eqs. \ref{lambda1} and \ref{lambda2} respectively. 

\subsubsection{Odd Hamilonians}
For generic Hamiltonians odd in $w$, Eq. \ref{RG3} nicely simplifies to
\begin{equation}
P(w)-1=\lambda\frac{h(w^2)}{h(w)}=\frac1{2^\gamma}\frac{h(w^2)}{h(w)}
\label{Podd}
\end{equation}
This equation can always be satisfied by 
\begin{equation}
h(w)=\prod_j \left(\frac{w^{a_j}-w^{-a_j}}{2i}\right)^{b_j}
\end{equation}
i.e. $h(k)=\prod_j \sin^{b_j}(a_jk)$ for $\sum_j a_jb_j \in \text{odd}$. From the familiar relation $\frac{\sin 2x}{2\sin x} =\cos x$, we see that $P(w)$ is a valid wavelet autocorrelation polynomial given by $P(w)=1+ \prod_j \left(\frac{w^{a_j}+w^{-a^j}}{2}\right)^{b_j}$. Two interesting special cases are elaborated below.

Hamiltonians of the form $h(k/2)=\sin \frac{ak}{2} = \frac{w^{a}-w^{-a}}{2i}$, $a$ odd, are invariant under the IR filter $C(w)=\frac{1+w^{a}}{\sqrt{2}}$ or $P(w)=1+ \frac{w^{a}+w^{-a}}{2}$, with $\lambda =\frac{1}{2}$. We need $a$ to be odd as $P(w)$ can never have even nontrivial even powers.

The above results are applicable to Hamiltonians even in $k$ too, as long as they are odd in $w=e^{ik/2}$, i.e. Hamiltonians of the form
\begin{equation}
h(k)=\cos ak-\cos bk\sim \frac{b^2-a^2}{2}\left(k^2+\frac{a^2+b^2}{12}k^4\right)
\end{equation}
since $h(k/2)= \frac{w^{a}+w^{-a}}{2}-\frac{w^{b}+w^{-b}}{2}$, $a,b$ odd, are invariant under $P(w)=C(w)C^*(w^{-1})=1+ \frac{w^{a}+w^{-a}+w^b+w^{-b}}{4}$, with $\lambda =\frac{1}{4}$. To find $C(w)$, note that $P(w)$ can always be factorized into $C(w)$ and $C^*(w^{-1})$ because it is symmetric in $w$ and $w^{-1}$, and its roots hence comes in pairs of $w$ and $w^{-1}$. This factorization admits no general analytic solution, but for simple cases like $a=3,b=1$, we can (with a bit of effort) find the nice solution 
$C(w)=\frac{1-i(w+w^2)+w^3}{2}$. This defines the wavelet basis for which $h(k)=\cos 3k -\cos k$ remains invariant.

With odd Hamiltonians, one can directly check from the form of $h(w)$ if the corresponding wavelet is of $\kappa=1$. Such bases are characterized by a nonvanishing $P''(1)$, which can be obtained via direct differentiation of Eq. \ref{Podd}:
\begin{equation}
 P''(1)=\frac{3h''(1)+2h'''(1)}{2h'(1)}-\frac1{2}\left(\frac{h''(1)}{h'(1)}\right)^2 \end{equation}
Evidently, some fine-tuning is needed to necessitate a wavelet with $\kappa>1$ (i.e $P''(1)=P'''(1)=0$).

\section{Wavelet dependence of bulk geometry }

One of the most attractive features of the Exact Holographic Mapping is that it reproduces, for various important cases, bulk geometries in agreement the Ryu-Takayanagi (RT) formula\cite{ryu2006}. Specifically, it yields for any number of dimensions the AdS space for critical systems at zero temperature, and BTZ/Lifshitz black holes for critical linear/nonlinear dispersing systems at nonzero temperature\cite{qi2013,lee2016exact}.

The RT formula proposes that the the entanglement entropy of a boundary region is proportional to the area of its corresponding minimal surface in the bulk. Inspired by this information theoretic\footnote{There has also been parallel studies on criticality based on information theory, c.f. Refs. \onlinecite{matsueda2012holographic,matsueda2013tensor,imura2014snapshot,lee2014exact,matsueda2014comment,matsueda2015proper,lee2016random,matsueda2016inverse}.} definition of area, the EHM framework proposed\cite{qi2013,lee2016exact} that geodesic distances in the EHM bulk are \emph{determined} by mutual information, i.e. the upper bound of the correlation functions between two endpoints. This is a paradigm shift from the usual conceptual relationship between correlation and distance: Conventionally, we think of the correlator decay behavior as a function of separation distance but now, we invert this relationship by \emph{defining} the distance based on the extent of correlator decay.

In this section, we shall focus on the the dependence of the bulk geometry on the wavelet basis, which is an aspect not studied in Ref. \onlinecite{lee2016exact}.

\subsection{Definition of the bulk geometry}
Consider two points $1$ and $2$ in the bulk system with coordinates $(\vec x_1,n_1,t)$ and $(\vec x_2,n_2,t)$, where $ \vec x$ is the site index within a level, $n$ the level index and $t$ the time. These two points are separated by a spatial coordinate interval of $\Delta \vec x=(2^{n_1}x_1-2^{n_2}x_2, n_1-n_2)$ sites and temporal coordinate interval of $\Delta t$. Recall that each level in the bulk contains $\propto 2^{-n}$ DOFs with spectral weight $|W_n(k)|^2$, such that we approach the low energy limit in the limit of large $n$. 

With the EHM, we \emph{define} the physical distance $d_{12}$ between these two points in the bulk by 
\begin{equation}
d_{12}=-\frac{d_0}{2}\log I_{12}\sim  -d_0\log C_{12}
\label{d}
\end{equation}
where $I_{12}$ is the mutual information between points $1$ and $2$ and $C_{12}$ is the two-point \emph{bulk} correlation function between them. The length scale $d_0$ can be interpreted as the inverse mass scale of the massive field associated with $C_{12}$ living in the \emph{curved} bulk geometry. The asymptotic equality on the RHS was shown in Ref. \onlinecite{lee2016exact}, that $I_{12}$ behaves asymptotically like $ 8C_{12}^2$. Eq. \ref{d} also applies for temporal intervals if we perform a Wick rotation to imaginary time $\tau=it$, so that temporal oscillations become exponential decay. With that, we have
\begin{align}
C_{12}(\Delta \vec x,\tau)&=\langle Tb_{n_2x_2}(\tau)b^\dagger_{n_1x_1}(0)\rangle\notag\\
&=\sum_k W^*_{n_1}(e^{-ik})W_{n_2}(e^{ik})e^{-ik(2^{n_1}x_1-2^{n_2}x_2)}G_{k}(\tau)\notag\\
&=\oint_{|z|=1}\frac{dz}{z}W^*_{n_1}(z^{-1})W_{n_2}(z)z^{2^{n_2}x_2-2^{n_1}x_1}G_z(\tau)
\label{correlator}
\end{align}
in terms of the \emph{boundary} correlation function $G_k(\tau)$ ($G_z$ and $G_k$ are used interchangeably, depending on the argument used) given by
\begin{equation}
G_k(\tau)=\frac{e^{\tau h(k)}}{\mathbb{I}+e^{\beta h(k)}}
\end{equation}
for the Hamiltonian $h(k)$, with $\beta$ the inverse temperature. Near a gapless point $z=e^{ik}=1$, the energy manifolds (eigenenergy bands of $h(z)$) generically exhibit branch points\footnote{The energy manifolds are the solutions to the characteristic polynomial associated with the eigenvalue equation.}. As we see later, the power-law decay of $C_{12}$ shall depend crucially on the existence of these complex singularities. In a typical case without accidental degenaracy, the band crossing involves two bands and $G_k$ possesses a square-root branch cut $u\sim \sqrt{z}=e^{ik/2}$ or $u\sim \sqrt{z^{-1}}=e^{-ik/2}$. To see this explicitly, consider the canonical two-band Dirac model $h(k)=\sin k\sigma_1 + (m+1-\cos k)\sigma_2$, $\sigma_{1,2}$ the Pauli matrices, with eigenenergies $E_k=E_z=\sqrt{1+(m+1)^2-(m+1)(z+\frac{1}{z})}$ and gap $m$. In matrix form,
\begin{equation}h(z)=\left(\begin{matrix}
 & 0 & i(\frac{1}{z}-(1+m)) \\
 & -i(z-(1+m)) & 0 \\
\end{matrix}\right)\label{hz}\end{equation}
with the correlator $G_z$ given by
\begin{eqnarray}
\left(\cosh(\tau E_z)\mathbb{I}+\frac{h(z)}{E_z}\sinh(\tau E_z)\right)\left(\mathbb{I}-\frac{h(z)}{E_z}\tanh\frac{\beta E_z}{2}\right)\notag\\
\label{gq}
\end{eqnarray}
Crucial to the analytic structure of this matrix is the "flattened hamiltonian" 
\begin{eqnarray}
\frac{h(z)}{E_z}&=&\left(\begin{matrix}
 & 0 & \sqrt{\frac{m+1}{z}}\sqrt{\frac{z-\frac{1}{m+1}}{z-(m+1)}} \\
 & \sqrt{\frac{z}{m+1}}\sqrt{\frac{z-(m+1)}{z-\frac{1}{m+1}}} & 0 \\
\end{matrix}\right)\notag\\
&\rightarrow_{m=0}& \left(\begin{matrix}
 & 0 & \frac1{\sqrt{z}} \\
 & \sqrt{z}& 0 \\
\end{matrix}\right)
\label{hz}\end{eqnarray}
Its branch cut topology crucially affects the bulk correlator because it dictates the deformation of the contour in Eq. \ref{correlator}. In the gapped case with nonzero $m$, $\frac{h(z)}{E_z}$ has 4 branch points $(0,\infty,m+1,\frac{1}{m+1})$, two within and two outside the unit circle. Hence $C_{12}$ can be evaluated without deforming the unit circle, giving rise to results\cite{lee2016exact,lee2016band} dependent on the position of the singularities introduced by either mass or temperature scale, but \emph{independent} of the wavelet basis.

In the gapless ($m=0$) case which we shall focus on, the only\footnote{No branch cut can be introduced by wavelet functions $W(z)$ and $W^*(z^{-1})$, which are polynomials.} branch cut extends from $z=0$ to $z=\infty$, which is unavoidable. In the following, we shall evaluate the bulk correlator and hence bulk geodesic distances by deforming the unit circle to a keyhole-like contour, from which the dependence of the correlator decay behavior on the branch cut becomes apparent. We shall consider the general case where the unitary transforms (and hence filters $C_j$ and $D_j$) at each iteration $j$ are not necessarily the same.

\subsection{Geodesic distances and bulk geometry for a critical 1D free fermion}
\subsubsection{Intra-level direction}

To explicitly demonstrate how the bulk geometry depend on the choice of wavelet basis, we turn to the simplest case of critical 1D free fermion described by a Dirac Hamiltonian. We stress that this choice of Hamiltonian is made purely due to its analytic tractability; indeed, an EHM generalized to arbitrary wavelet bases will be able to retain the forms of a far larger class of Hamiltonians (Sect. III). 

We first study the zero-temperature bulk correlator $C_{12}$ due to a displacement of $x$ sites in the intra-level direction, so that level indices $n_1=n_2=n$ are equal and $\tau=0$, $\beta\rightarrow\infty$. Physically, this correlator is between degrees of freedom at the same scale and time.

A nonzero matrix element $u$ of $C_{12}$ is given by
\begin{eqnarray} u&=&-\oint_{|z|=1}\frac{dz}{z}W^*_{n}(z^{-1})W_{n}(z)\frac{\sqrt{z}}{2}z^{2^n x}\notag\\
&=&\frac{1}{2}\int_0^1 W_n^*(z^{-1})W_n(z)z^{2^nx}(\sqrt{z}-\sqrt{e^{2\pi i}z})\frac{dz}{z}\notag\\
&=&\int_0^1 W_n^*(z^{-1})W_n(z)z^{2^nx}\frac{1}{\sqrt{z}}dz\notag\\
&=&\int_0^1 (W_n^*(z^{-1})W_n(z)z^{m2^n})z^{2^n(x-m)-1/2}dz\notag\\
&=&\int_0^1 Q(z)z^X dz\notag\\
&=&\frac{Q(1)}{X+1}-\frac{1}{X+1}\int_0^1Q'(z)z^{X+1}dz\notag\\
&=& \sum_{j=0} \frac{Q^{(j)}(1)X!}{(X+j)!}(-1)^j\notag\\
&\sim& \frac{Q^{(2\kappa )}(1)}{X^{2\kappa +1}}\notag\\
&\sim& \frac{Q^{(2\kappa )}(1)}{2^{n(2\kappa +1)}x^{2\kappa +1}}
\label{u2}
\end{eqnarray}
In line 4, $m$ is the degree of each factor $C$ or $D$ in $W_{n}(z)=\frac{1}{2\pi}D_n(z^{2^{n-1}})\prod_{j=1}^{n-1}C_j(z^{2^{j-1}})$, introduced such that $Q(z)=W_n^*(z^{-1})W_n(z)z^{m2^n}$ does not have negative powers of $z$. In line 5, $X=2^n(x-m)-1/2$ is large and positive for fairly large intervals $x$, so that the $j$ corrections in the third last line can be dropped. The final expression involves $\kappa$, the first nonzero derivative of $W_n(z)$ at $z=1$ (see Sect. \ref{sec:holowav}).  
The integer $\kappa$, which characterizes the wavelet moment at the IR (long-wavelength) point $z=1$, shall be a key quantity in determining how the EHM affects the correlators and hence bulk geometry.

Let's now evaluate $Q^{(2\kappa )}(1)$ by an explicit expansion about $z=1$:
\begin{eqnarray}
&&\frac{Q^{(2\kappa )}(1)\epsilon^{2\kappa }}{(2\kappa )!}\notag\\
&=& Q(1-\epsilon)\notag\\
&=& (1-\epsilon)^{2^nm}W_n^*((1-\epsilon)^{-1})W_n(1-\epsilon)\notag\\
&\approx&W_n^*(1+\epsilon)W_n(1-\epsilon)\notag\\
&=&\frac{1}{2\pi} \left |\left(\prod_{j=1}^{n-1}C_j\left((1-\epsilon)^{2^{j-1}}\right)\right)D_n\left((1-\epsilon)^{2^{n-1}}\right)\right|^2\notag\\
&\approx &\frac{1}{2\pi} \left |\left(\prod_{j=1}^{n-1}C_j(1)\right)D_n\left(1-2^{n-1}\epsilon\right)\right|^2\notag\\
&\approx &\frac{1}{2\pi} \left |\left(\prod_{j=1}^{n-1}C_j(1)\right)\right|^2\left |D_n^{(\kappa )}(1)\frac{2^{\kappa (n-1)}\epsilon^\kappa }{\kappa !}\right|^2\notag\\
&\approx &\frac{1}{2\pi} \left |\left(\prod_{j=1}^{n-1}C_j(1)\right)\right|^2\left |D_n^{(\kappa )}(1)\right|^2\frac{2^{2\kappa (n-1)}\epsilon^{2\kappa }}{\kappa !^2}
\label{u4}
\end{eqnarray} 
The $C$ factors are evaluated at $z=1$ with no need for Taylor expansion because they are IR filters, which are not supposed to have vanishing values at $z=1$. Comparing coefficients, we see that 
\begin{eqnarray}
Q^{(2\kappa )}(1)&=& \frac{1}{2\pi} \left |\left(\prod_{j=1}^{n-1}C_j(1)\right)\right|^2\left |D_n^{(\kappa )}(1)\right|^22^{2\kappa (n-1)}\left(\begin{matrix}
 &2\kappa  \\ &\kappa   \\ \end{matrix}\right)\notag\\
\label{u5}
\end{eqnarray}
$D_n^{(\kappa )}(1)$ is the first nonzero derivative of the UV filter $D_n$ at the IR point $k=0$ or $z=1$. Combining Eq. \ref{u5} with Eqs. \ref{d} and \ref{u2}, we obtain 
\begin{eqnarray}
I_{12}&\sim & 8u^2\notag\\
&=& \left(\left(\begin{matrix}
 &2\kappa  \\ &\kappa   \\ \end{matrix}\right)\frac{|D_n^{(\kappa )}(1)|^2}{2^{2\kappa }\pi}\right)^2\left |\left(\prod_{j=1}^{n-1}\frac{C_j(1)}{\sqrt{2}}\right)\right|^2\frac{1}{2x^{4\kappa +2}}\notag\\
&=& \left(\begin{matrix}&2\kappa  \\ &\kappa   \\ \end{matrix}\right)^2\frac{|D_n^{(\kappa )}(1)|^4}{2^{4\kappa+1 }\pi^2} \prod_{j=1}^{n-1}\left(1-\frac{|C_j(-1)|^2}{2}\right)\frac{1}{x^{4\kappa +2}},\notag\\
\label{Ixy3}
\end{eqnarray}
which coincides with results from Ref. \onlinecite{singh2016holographic} for \emph{bosonic} systems. All in all we have (plotted in Fig. \ref{fig:corr})
\begin{equation}
d_{12}(x)\sim d_0(2\kappa +1)\log |x| +\text{const.}
\label{I12x}
\end{equation}
Explicitly, we see that the mutual information $I_{12}$ decays with $x$ with an exponent of $4\kappa+2$, i.e. that the choice of wavelet basis affects the coefficient $4\kappa+2$ of the logarithmic term, but does not modify its qualitative asymptotic behavior. Physically, a larger $\kappa$ leads to faster decay of mutual information because the additional smoothness of the UV filter $D_n$ at $k=0$ extinguishes more DOFs. Notably, there will be no dependence on $n$, the level index, only if $C_j(-1)=C_j(k=\pi)=0$ for all levels $j$. In other words, each IR filter $C_j$ will lead to a suppression of $I_{xy}$ unless $C_j(-1)=0$, i.e. is a perfect IR filter taking zero value at the UV point $k=\pi$. To put the significance of this observation in context, consider the fitting of the geodesic distance $d_{12}\sim d_0 \log C_{12}$ with that of Anti de-Sitter (AdS) space (Appendix I of Ref. \onlinecite{lee2016exact}):
\begin{equation}
d_{AdS}(x)\sim 2R\log \frac{|x|}{R} 
\label{I12xA}
\end{equation}
where $R$ is the AdS radius. If we want to fit $d_{12}(x)$ of $I_{12}$ to $d_{Ads}(x)$, which do not depend on the radial coordinate, we will need each iteration of the EHM to discard \emph{all} of the largest scale DOFs, which are at $k=\pi$. This can only happen if $C_j(-1)=0$ for all levels $j$. Merely having all $C_j$'s equal is not sufficient for ensuring that the geodesic distance is independent of the scale $n$.

From now, we assume perfect IR filters that have zero support at $k=\pi$. Comparing Eqs. \ref{I12x} and \ref{I12xA}, we obtain{
\begin{equation}
\frac{R}{d_0}=\kappa +\frac1{2}
\label{Ixy4}
\end{equation}
and
\begin{equation}
R=\frac{1}{2}\left(\frac{\sqrt{2}|D_n^{(\kappa )}(1)|^2\left(\begin{matrix}
 &2\kappa  \\ &\kappa   \\ \end{matrix}\right)}{\pi}\right)^{\frac{1}{2\kappa +1}}
\label{Ixy5}
\end{equation}
We see that $R$ depends only on $\kappa $ and $|D_n^{(\kappa )}(1)|$. In the simplest case of the Haar wavelet basis, $D_n(z)=\frac{1-z}{2}$, so $\kappa =1$ and $|D_n^{(1)}(1)|=\frac{1}{\sqrt{2}}$. Eq. \ref{Ixy4} and \ref{Ixy5} then coincides with numerical results from Ref. \onlinecite{qi2013}.  

\begin{figure}[H]
\includegraphics[scale=.75]{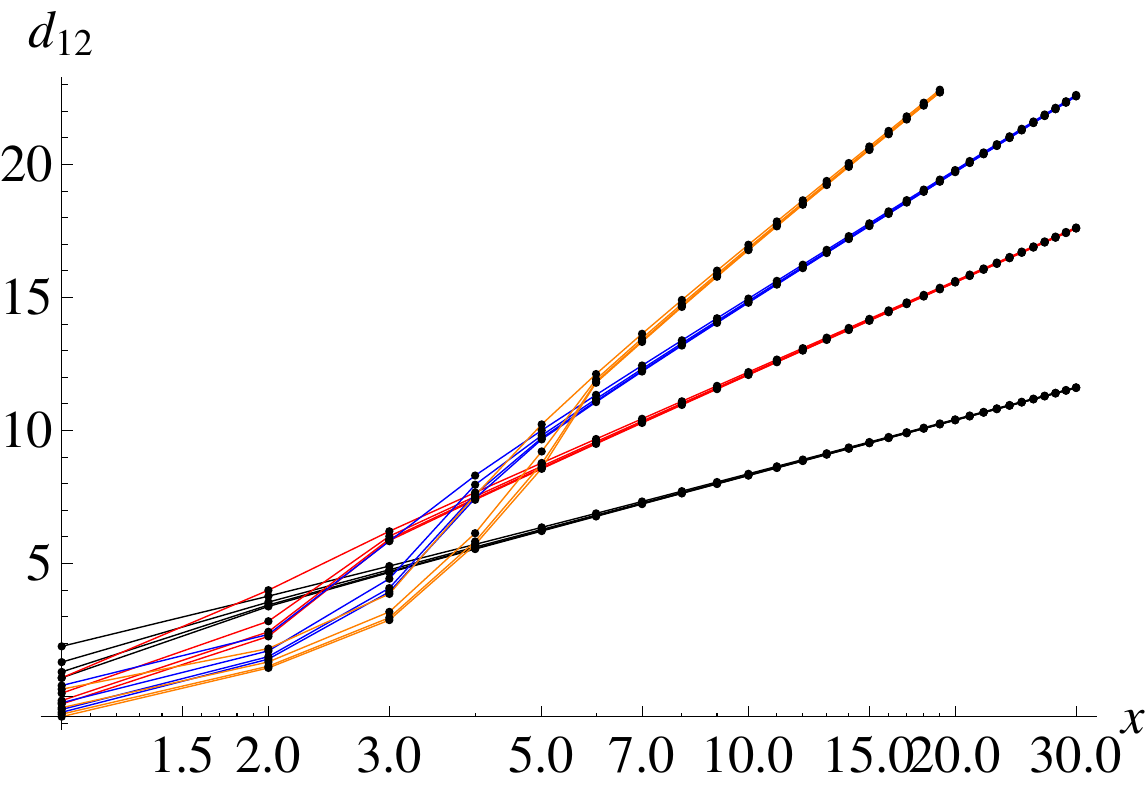}
\caption{The geodesic distances $d_{12}$ determined from the bulk correlators via Eq. \ref{d}. Plotted are the curves for the Dirac model with the $\kappa=1,2,3$ and $4$ Daubechies wavelet filters. Their excellent numerical agreement with Eq. \ref{Ixy4} or \ref{Ixy5} (continuous straight lines) show that the AdS radius is indeed directly proportional to $\kappa$, i.e. $2\kappa+1=3,5,7$ and $9$ respectively. That $d_{12}$ exhibits power-law decay after just a few sites provide a posteriori justification of the approximations in Eq. \ref{u2}.}
\label{fig:corr}
\end{figure}

\subsubsection{Inter-level (radial) direction}

We now consider the case with zero intra-level displacement ($x=0$) and temporal displacement ($\tau =0$), so that the interval lies in the "adial" direction from level $1$ to level $n$. This is an interval between different lengths scales at the same spacetime coordinates. A nontrivial matrix element $u$ of the bulk correlator takes the form
\begin{equation} u=-\oint_{|z|=1}\frac{dz}{z}W^*_{1}(z^{-1})W_{n}(z)\frac{\sqrt{z}}{2}
\label{u}
\end{equation}
This is a complicated expression that admits no general simplification. However, its asymptotic behavior can be computed as follows: Define
\begin{equation}
I_n=u=-\frac1{2}\oint_{|z|=1}\frac{dz}{z}W^*_{1}(z^{-1})D_n(z^{2^{n-1}})\prod_{j=1}^{n-1}C_j(z^{2^{j-1}})\sqrt{z}
\end{equation}
and 
\begin{equation}
J_{n-1}=-\frac1{2}\oint_{|z|=1}\frac{dz}{z}W^*_{1}(z^{-1})\prod_{j=1}^{n-1}C_j(z^{2^{j-1}})\sqrt{z}
\end{equation}
which is the unprojected correlator in the $(n-1)^{th}$ level. For $2^n\gg 1$, $I_n$ and $J_{n-1}$ are approximately related by
\begin{eqnarray}
&&I_n\notag\\
&=& -\frac1{2}\oint_{|z|=1}\frac{dz}{\sqrt{z}}W^*_{1}(z^{-1})\prod_{j=1}^{n-1}C_j(z^{2^{j-1}})(D_n(0)+O(z^{2^{n-1}}))\notag\\
&\sim& D_n(0)J_{n-1}
\end{eqnarray}
since the truncated contributions from monomials of $z^{2^{n-1}}$ integrate to small quantities $\frac{1}{2^{n-1}+\text{const.}}$ that can be discarded for $2^n\gg 1$. Hence $I_n$ is dominated by the term containing $D_n(0)$, the constant term in $D_n(z)$. Note that $|D_n(0)|=|d(0)|<1$ since $\sum_j |d(j)|^2=1$. Similarly, we can also show that $J_n\sim C_n(0)J_{n-1}$. Hence asymptotically,
\begin{eqnarray} 
|u|\sim D_n(0)\prod_{j=1}^{n-1}C_j(0)&\propto & C(0)^{n-1},
\label{radial}
\end{eqnarray} 
the last expression holding when the IR filters $C_j$ are all the same. Hence the radial geodesic distance goes like
\begin{equation} 
d_{12}(1,n)= -d_0\log |u|  \sim (n-1)\log\frac{1}{C(0)^2}+\text{small const.}
\label{radial2}
\end{equation} 
Comparing this with the radial AdS distance\cite{qi2013,lee2016exact}
\begin{equation} 
d_{AdS}(1,n)\sim R(n-1)\log 2,
\label{radial2A}
\end{equation}
we obtain
\begin{equation}
\frac{R}{d_0}= -\frac{2\log |C(0)|}{\log 2} 
\label{Rradial}
\end{equation}
so that $\frac{R}{d_0}=1$ in the Haar case with $C(z)=\frac{1+z}{\sqrt{2}}$. 

$C(0)$ is the same-site coefficient in the real-space recursion relation of the IR wavelet filter. As such, a small $C(0)$ represents a large 'spreading' of the EHM tree network, and should cause the mutual information to decay faster as we travel down the different hierarchical levels $(n)$ of the the tree.


\subsubsection{Imaginary time direction}

We now focus on the case with $\Delta \vec x =0$, but imaginary time interval $\tau>0$. From Eq. \ref{gq}, the leading contribution to the correlator is 
\begin{equation}
C_{12}(\tau)=\frac{1}{2}\int_{-\pi}^\pi dq |W_n(e^{iq})|^2 e^{-\tau E_q}
\label{time}
\end{equation}
The simplying caveat is that we only have to care about the extreme IR (small $q$) contribution to this integral. This is because $e^{-E_q\tau}=e^{-v_F|q|\tau}$ decays rapidly for moderately large $\tau$. Hence we only need to know the IR behavior of $W_n(z)=D_n(z^{2^{n-1}})\prod_{j=1}^{n-1}C_j(z^{2^{j-1}})$, which is given by Eq. \ref{u4}:
\begin{eqnarray}
|W_n(e^{i \Delta q})|^2&\approx & |W_n(1-i \Delta q)|^2\notag\\
&\approx & \left |\left(\prod_{j=1}^{n-1}C_j(1)\right)\right|^2\left |D_n^{(\kappa )}(1)\right|^2\frac{2^{2\kappa (n-1)}(\Delta q)^{2\kappa }}{2\pi\kappa !^2}\notag\\
&=&\frac{1}{2\pi} \left |D_n^{(\kappa )}(1)\right|^2\frac{2^{(2\kappa +1)(n-1)}(\Delta q)^{2\kappa }}{\kappa !^2}
\label{u6}
\end{eqnarray}
where $\kappa $ is the order of the first nonzero derivative of $D_n(z)$, as before. Evaluating Eq. \ref{time} in terms of the incomplete Gamma function $\int_0q^\gamma e^{-q\tau}dq\sim\frac{\gamma!}{\tau^{\gamma+1}}$, we obtain
\begin{eqnarray}
C_{12}(\tau)&\sim&\frac{1}{2 \pi} \left |D_n^{(\kappa )}(1)\right|^2\frac{2^{(2\kappa +1)(n-1)}}{\tau^{2\kappa +1}}\left(\begin{matrix}
 &2\kappa  \\ &\kappa   \\ \end{matrix}\right)
\label{time3}
\end{eqnarray}
Comparing $d_{12}(\tau)=-d_0\log C_{12}(\tau)$ with the imaginary time geodesic distance of Euclidean AdS space\cite{qi2013,lee2016exact}
\begin{equation}
d_{AdS}(\tau)=2R\left(\log \frac{\tau}{R}-n\log 2\right),
\end{equation}
we obtain
\begin{equation} \frac{R}{d_0}=\kappa +\frac{1}{2} \end{equation}
which agrees exactly with the intra-level result (Eq. \ref{Ixy4}). The corresponding AdS radius $R$ is also given by Eq. \ref{Ixy5}. 

The equivalence of the fitting parameters to AdS space in the intra-level and imaginary time directions is not surprising, since there is a global rotation symmetry that relates space and imaginary time.

\section{Further generalizations}
\subsection{``Zooming in'' onto arbitary Fermi points}
\label{sec:zoomin}

The EHM is essentially a ``lossless'' RG procedure producing a series of bulk layers $n$ that represent the original system viewed from various energy scales. Mathematically, that is accomplished by ``zooming in'' successively closer to the low energy regions of the system. In a fermionic system, the lowest energy regions are Fermi points in the case of semimetals, or Fermi surfaces in the case of metals. It is imperative that we are not just able to probe the long-wavelength $k=0$ limit, but also able to probe the low energy limit of a given system. Since the EHM should fundamentally be a low energy probe, the resultant bulk geometry should not be qualitatively affected by that positions of the Fermi points. That this is true will be evident from the results of this section, where we show that all that is required is a modification of the wavelet basis.

So far, the EHM described involve iterations that successively ``zoom in'' onto the long wavelength limit $k=0$ (or $z=e^{ik}=1$). This is appropriate if the physical system has a Fermi point at $k=0$. However, most real systems like Graphene\cite{allen2009honeycomb} or specially design metamaterials\cite{sun2012topological,gao2016classification,lin2016dirac,lee2015negative,yan2016tunable,muechler2016tilted,yan2017nodal,yan2017experimental,bi2017nodal,li20172,li2017engineering} possess interesting and possibly topologically nontrivial\footnote{When there is a continuum of Fermi points that form an extended Fermi surface, we require a different type of EHM involving conformal maps (work in progress).} critical points (valleys, line nodes etc.) elsewhere in the Brillouin zone. 

If the critical point is simply shifted to $k_0\neq 0$, we can trivially modify the EHM via 
\begin{equation}
C(z)\rightarrow C(ze^{-ik_0}), \;\;\; D(z)\rightarrow D(ze^{-ik_0})
\label{wavelet30}
\end{equation}
so that its spectral properties are simply translated by $k_0$. This modification introduces complex coefficients in the real-space wavelet functions, which is perfectly permissible for a wavelet mapping acting in quantum mechanical Hilbert space.

More interestingly, we can also ``split'' the spectral peaks such that the EHM ``zooms in'' onto more than one momentum point. This is achieved by interchanging the sequence of UV and IR filters in the tower of $C$ and $D$ filters used in constructing  $W_n(z)$ in Eq. \ref{wavelet2}: Instead of the original definition $W_n(z)=C(z)C(z^2)...C(z^{2^{n-2}})D(z^{2^{n-1}})$, we shall define
\begin{eqnarray}
W_n(z)&=& B^+_n\left(z^{2^{n-1}}\right)\prod_j^{n-1}B^-_j \left(z^{2^{j-1}}\right)
\label{wavelet3}
\end{eqnarray} 
where each $B_n^\pm(z)$ can be \emph{either} $C(z)$ or $D(z)$. We define a vector $\vec v$ such that $v_j=1$ if $C(z)$ was used at level $j$, and $v_j=-1$ if $D(z)$ was used. Hence the usual definition of $W_n(z)$ will correspond to $\vec v=(1,1,1,...,1,-1)$, while $C(z)D(z^2)D(z^4)C(z^8)D(z^{16})$, for instance, will correspond to $\vec v=(1,-1,-1,1,-1)$. 

The effects of interchanging the $C$ and $D$ filters are illustrated in Fig. \ref{fig:CD}. At each level, the IR filter $C\left(z^{2^n}\right)$ has vanishing spectral weight when $z^{2^n}=-1$, i.e. $k=\frac{\pi}{2^{n-1}}(2j+1)$, $j\in \mathbb{Z}$. If the tower of filters take the form $C(z)C(z^2)C(z^4)...$, i.e. consists of all IR filters $C$, $k=0$ eventually survives as the only peak. In this sense, $W_n(z)$ zooms in onto $k=0$. 

\begin{figure}[H]
\includegraphics[scale=.5]{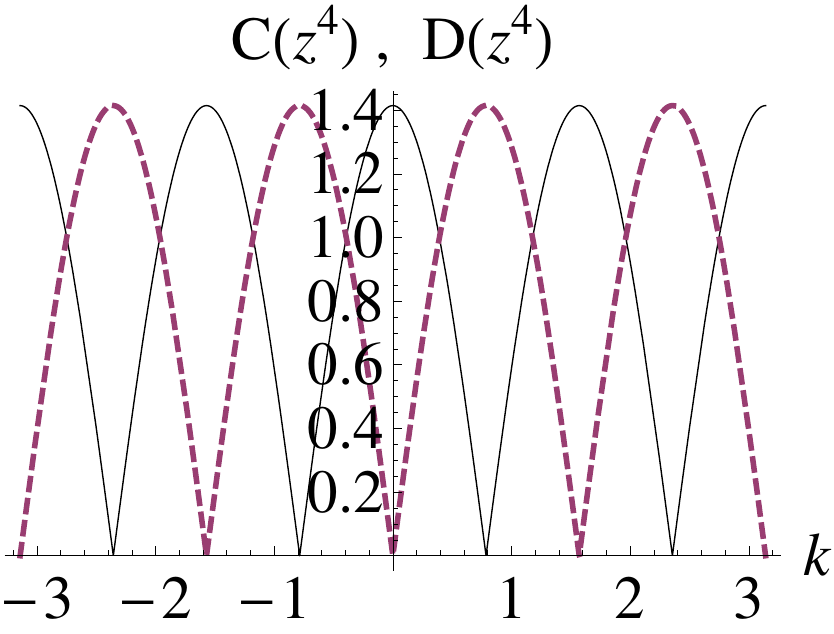}
\includegraphics[scale=.5]{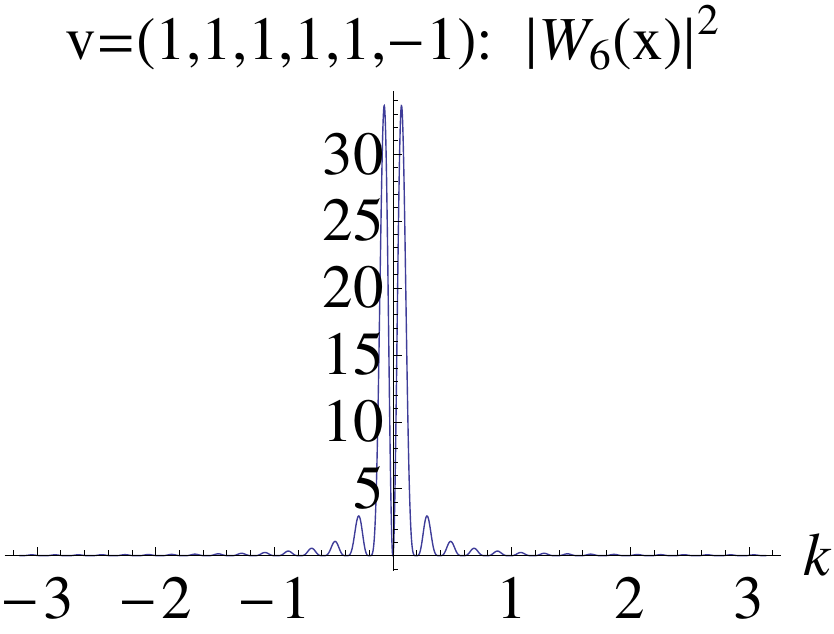}
\includegraphics[scale=.5]{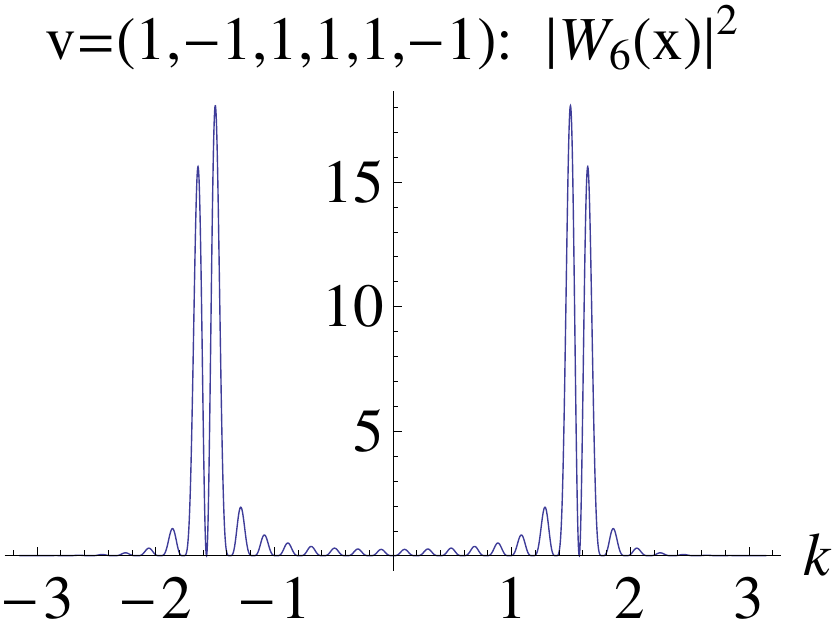}
\includegraphics[scale=.5]{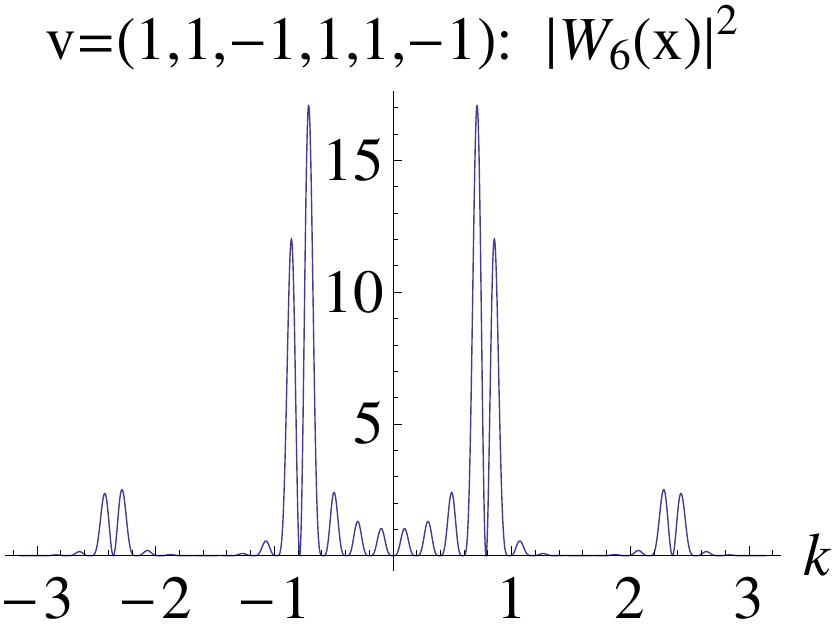}
\caption{Top Left) Illustration of the profiles of $C$ (black) and $D$ (purple dashed) for the Haar wavelet. The IR filter $C\left(z^{2^n}\right)$ strengthens the (IR) contribution closer to the existing folded IR points at $\frac{2\pi}{2^n}j$, $j\in \mathbb{Z}$. The UV filter $D\left(z^{2^n}\right)$  attempts to ``split'' these contributions by favoring contributions halfway between the IR points. Top Right) $|W_6(z)|^2$ as defined in the usual case of $CCCCCD$. Each $C$ filter strengthens the peak around $k=0$, while the last $D$ filter splits it to support the contributions just around $k=0$. Bottom Left) A given $D$ filter at the second level splits the peaks to $\pm \pi/2$. Bottom Right) A $D$ filter at the third level splits the peaks to $\pm \pi/4$. The secondary peaks will be much more attenuated if $\kappa>1$ was used. }
\label{fig:CD}
\end{figure}

Now, suppose that $C\left(z^{2^m}\right)$ is replaced by $D\left(z^{2^m}\right)$, which suppresses $k=\frac{2\pi}{2^{m-1}}j$, $j\in \mathbb{Z}$. This includes the $k=0$ point, which will thus no longer be ``zoomed in'' onto. But at the same time, the points $k=\frac{\pi}{2^{m-1}}(2j+1)$, $j\in \mathbb{Z}$ will be allowed to survive. Due to the finite envelope of $C$ and $D$ as shown in Fig. \ref{fig:CD}, the spectral weights of these new peaks depends on the distance from the previous IR point. Consider the example $\vec v= (1,-1,1,1,1,-1)$. At $m_1=2$, $C(z^2)$ is replaced by $D(z^2)$. This replaces the ``default'' IR peak of $k=0$ by the ``new'' IR peaks $\pm \frac{\pi}{2}$. At the next $m_2=6$, $C(z^{32})$ is replaced by $D(z^{32})$. Of all the k-points $k=\frac{\pi}{32}(2j+1)$, $j\in \mathbb{Z}$, that thus do not have to vanish, the dominant ones are those closest to the incumbent IR points $\pm \frac{\pi}{2}$. In general, when $C$ is replaced by $D$ at $z^{2^{m_1}}, z^{2^{m_2}},...,z^{2^{m_r}}$, the dominant pair of IR points eventually zoomed in onto will be $\pm \frac{\pi}{2^{m_1-1}}\pm  \frac{\pi}{2^{m_2-1}}\pm ...\pm \frac{\pi}{2^{m_r-2}}$, where the $j^{th}$ $\pm$ sign ($j> 2$) is chosen such that the point to be zoomed in is closer to the ``old'' IR point $\pm \frac{\pi}{2^{m_1-1}}\pm ...\pm\frac{\pi}{2^{m_{j-2}-1}}$.

To zoom in onto arbitrary Fermi points, one can combine translations and splittings of the IR points at various levels via Eqs. \ref{wavelet30} and \ref{wavelet3}, as well as utilize specific forms of $C_j$ and $D_j$ to achieve the desired spectral peaks.

\subsection{EHM in higher dimensions - basis anisotropy}

While we have so far focused on 1-dimensional EHM, all the results so far can be directly generalized to a higher dimensional EHM relating a $d+1$-dim boundary system to a $d+2$-dim bulk. This generalization can be simply accomplished by taking direct products of the wavelet filters in various dimensions, as described at length in Section V of Ref. \onlinecite{lee2016exact}. With possibly different $\kappa$ parameters $\kappa_1,...,\kappa_d$ for the wavelet basis in each direction, one may naively think that we will arrive at bulk geodesic distances given by $d_{12,j}(x_j)\sim d_0(2\kappa_j+1)\log x_j
$ with anisotropic AdS radii $R_j=d_0\left(\kappa+\frac1{2}\right)$. This is actually not true. To understand why, note that each factor of $|W_n(e^{ik})|^2$ near $k=0$ acts as a derivative on the (original) boundary correlator
\begin{equation}
G_x\propto \int G_k e^{ikx}\sim \frac1{|\vec x|}=\frac1{\sqrt{x_1^2+...+x_d^2}},
\end{equation}
so that for a wavelet filter in the direction $j$, $\int |W_n(e^{ik_j})|^2 e^{i2^n\vec k\cdot \vec x}G_k dk_j\sim \partial^{2\kappa_j}_j\frac1{|\vec x|}\sim \frac{3\cos^2\theta-1}{|\vec x|^3}$ where $\cos\theta$ is the $d^{th}$ component ratio of $\vec x$. 

In general, the bulk correlator $C_{12}(\vec x)$ will be dominated by terms involving the lowest $\kappa_j$ in almost \emph{all} directions, not just in the $j^{th}$ direction. To see why, consider the $d=2$ case $\kappa_1=1$ and $\kappa_2=2$. The two leading contributions to $C_{12}(\vec x)$ are proportional to
\begin{equation}
\partial^2_1\frac1{|\vec x|} = \frac{3\cos^2\theta-1}{|\vec x|^3}
\end{equation}
and 
\begin{equation}
\partial^4_2\frac1{|\vec x|} = \frac{3(3\cos^4\theta-24\cos^2\theta\sin^2\theta+8\sin^2\theta)}{|\vec x|^5}
\end{equation}
In the asymptotic limit of large $|\vec x|$, the decay exponent is \emph{always} $3$ unless $3\cos^2\theta-1$ is exactly zero, which is an interval of measure zero. Hence in the multidimensional case, we the AdS radius is generically given by $d_0\text{ min}(\kappa_1,...,\kappa_d)+d_0/2$.

\section{Conclusion}
\label{sec:concl}

Motivated by the desire for a holographic mapping that preserves the form of a wide class of Hamiltonians, we generalized the Exact Holographic Mapping to consist of the most general unitary transformation based on biorthogonal wavelets. Compared to the original EHM based on the Haar wavelet, our generalized EHM can preserve Hamiltonians with various exotic band touchings, and not just those of linear Dirac type. The precise relationship between the Hamiltonian and the wavelet mapping that preserves it is summarized in Eq. \ref{RG3}, which can also be shown to determine the renormalization scale factor $\lambda$. 

We also derived the dependence of the bulk geometry on the wavelet basis, and showed that the latter only affects quantities arising from branch cuts in the propagator. These include the correlator decay exponent of a critical system at zero temperature and hence its dual AdS radius, but not the spatial event horizon of the dual geometry due to mass or temperature scale. Of primary significance is the integer $\kappa$, which is the order of the first nonzero derivative of the IR wavelet filter $C(z)$ in the long-wavelength limit. It is $\kappa$, and not the length $2l$ of the mother wavelet, that controls the bulk geometry.

The generality of the wavelet EHM formulation also enables us to ``zoom in'' onto Fermi points away from the long wavelength limit. This can be accomplished, for instance, by reversing the roles of the UV and IR filters at certain scale levels $n$. Finally, we discussed the implications of having higher dimensional EHM with anisotropic bases.

We also took this opportunity to provide a pedagogical introduction to the construction of wavelets, a topic intimately related to renormalization but rarely covered in detail in the physics literature.

\begin{acknowledgements}
CH thanks Xiao-Liang Qi, Guifre Vidal and Yingfei Gu for helpful discussions. 
\end{acknowledgements}


\appendix

\section{Biorthogonality of wavelets}
\label{sec:biortho}

Consider the generic definition with the roles of the $C$ and $D$ wavelet filters possibly interchanged (Sect. \ref{sec:zoomin}). Eq. \ref{wavelet2} is generalized to
\begin{eqnarray}
W_n(z)&=& B^+_n(z^{2^{n-1}})\prod_j^{n-1}B^-_j (z^{2^{j-1}})
\end{eqnarray} 
where $B^\pm$ is the z-transform of $b^\pm$. It is possible to prove the orthogonality of the $W_n$'s from Eq. \ref{wavelet3}. Suppose $m>n$:
\begin{widetext}
\begin{eqnarray}
(w_n,w_m)&\propto &\oint_{|z|=1}\frac{dz}{z} W^*_n(z^{-1})W_m(z)\notag\\
&=&\oint_{|z|=1}\frac{dz}{z} B^{+*}(z^{-2^{n-1}})B^+(z^{2^{m-1}})\prod_{a=0}^{n-2}B^{-*}(z^{-2^a})\prod_{b=0}^{m-2}B^-(z^{2^b})\notag\\
&=&\oint_{|z|=1}\frac{dz}{z} \left[B^{+*}(z^{-2^{n-1}})B^-(z^{2^{n-1}})\right]\prod_{a=0}^{n-2} \left[B^{-*}(z^{-2^a})B^-(z^{2^a})\right]\left(1+O(z^{2^{n}})\right)\notag\\
&=& \oint_{|z|=1}\frac{dz}{z} (\text{nonconstant})\notag\\
&=&0
\label{ortho3}
\end{eqnarray} 
\end{widetext}
The first term in line 3, which is equal to $C^{*}(z^{-2^{n-1}})D^-(z^{2^{n-1}})$ or $D^{*}(z^{-2^{n-1}})C^-(z^{2^{n-1}})$, has no constant term by Eq. \ref{ortho1}, and has a smallest power of $\pm 2^{n-1}r$, $r$ an odd positive integer. This power cannot be canceled by any combination of terms in the product in the second term, since each term is equal to $C^{*}(z^{-2^a})C^-(z^{2^a})$ or $D^{*}(z^{-2^a})D^-(z^{2^a})$ and has no even power of $z^{\pm 2^a}$. Explicitly, each postive power term in the product has the form

\[ z^{\sum_{j=0}^{n-2}2^j(2m_j+1)}=z^{\sum_{j=1}^{n-1}m_{j-1}2^j+2^{n-1}-1}\]

where $m_j$ is either a non-negative integer or $-\frac{1}{2}$, the latter corresponding to the case when $z^{2^j}$ is not used. The exponent is thus odd and unable to cancel the power in $z^{-2^{n-1}r}$. This holds for the negative power terms too. The remaining terms from $W_m$ are either constant or have degree exceeding $\pm 2^{n-1}$, and so cannot form a constant term. Hence the integral is zero by the residue theorem.

If $w_n$ or $w_m$ were to be displaced from each other by a distance $x$, there will be an addition factor of $z^{\pm2^mx}$ or $z^{\pm2^nx}$ in the integral. However, it is clear from the above argument that such a term also cannot be combined with an other term to produce a constant term. Hence the displaced wavelet bases are also orthogonal, as required earlier on. 

Note that this above proof does not require $C$ and $D$ to be the same for each level $j$, but only that they must all satisfy the conditions mentioned in Sect. II. 




\section{Discussion on finding RG-invariant Hamiltonians}
\label{sec:matrix}
Here we give a matrix approach to solving for the $P(z)$ of the appropriate wavelet transform that leaves a given Hamiltonian $h(z)$ invariant.

General real Laurent polynomials $h(z)$ and $P(z)$ for $z=e^{ik}$, $k\in \mathbb{R}$ can be written as
\begin{equation}
h(z)=\sum^l_{j=0}a_jz^j+c.c.=h_{even}(z)+h_{odd}(z)
\end{equation}                                             
\begin{equation}
P(z)=1+ \left(\sum^l_{j\text{ odd}}p_jz^j+c.c.\right)
\end{equation}                                             
For $h(z)$ to be invariant under the wavelet transform described by $P(z)$ Eq. \ref{RG2},
\[\lambda h(z^2)=h_{even}(z)+(P(z)-1)h_{odd}(z)\]
must be satisfied. By equating the coefficients of non-negative powers of $z$ on both sides (there are only even powers, of course), we obtain the relation 
\begin{widetext}
\begin{equation}
\left(\begin{matrix}
 & a_l & 0 & 0 & ... & 0 & 0\\
 & a_{l-2} & a_l & 0& ... & 0 & 0\\
 & a_{l-4} & a_{l-2} &  a_ l & ... & 0 & 0 \\
& \vdots & \vdots & \vdots & \ddots & \vdots & \vdots \\
& a^*_{l-2} & a^*_{l-4} & a^*_{l-6} & ... & a_l & 0 \\
& a^*_l & a^*_{l-2} & a^*_{l-4} & ... & a_{l-2} & a_l
\end{matrix}\right) \left(\begin{matrix}
& p_l \\
& p_{l-2} \\
& p_{l-4} \\
& \vdots \\
& p^*_{l-2}\\
& p^*_l 
\end{matrix}\right) = \lambda\left(\begin{matrix}
& a_l \\
& a_{l-1} \\
& a_{l-2} \\
& \vdots \\
& a_1\\
& a_0
\end{matrix}\right)-\left(\begin{matrix}
& 0 \\
& 0 \\
& 0 \\
& \vdots \\
& a_2\\
& a_0
\end{matrix}\right)
\label{matrix1}
\end{equation}
\end{widetext}
where $l$ is the (odd) degree of $P(z)$, which is also the maximum possible degree of $h(z)$. In matrix equation form, Eq. \ref{matrix1} becomes $A\vec p = \lambda \vec a - \vec a_e$, where $A$ is the lower triangular Toeplitz matrix comprising the coefficients of $h_{odd}(z)$ and their complex conjugates, $\vec p$ and $\vec a$ the vectors of coefficients of $P(z)$ and $h_{odd}(z)$ respectively, and $\vec a_e$ the vector of $h_{even}(z)$. 

Fortuitously, the lower triangular matrix $A$ can be inverted easily. Writing $A=a_l ( \mathbb{I}+N)$ where $N$ is a Nilpotent Toeplitz matrix, we easily find that $A^{-1}=(\mathbb{I}-N + N^2 -...+ N^{l-1})/a_l$. Upon a bit more algebra, we find that
\begin{equation}
\left(\begin{matrix}
& p_l \\
& p_{l-2} \\
& p_{l-4} \\
& \vdots \\
& p^*_{l-2}\\
& p^*_l 
\end{matrix}\right) = \left(\begin{matrix}
 & b_0 & 0  & ... &0 & 0 & 0\\
 & b_1 & b_0 & ... &0 & 0 & 0\\
& \vdots & \vdots & \ddots & \vdots & \vdots & \vdots \\
& b_{l-2} & b_{l-3} & ... & b_0 & 0 & 0\\
& b_{l-1} & b_{l-2} & ... & b_1 & b_0 & 0\\
& b_{l} & b_{l-1} & ... & b_2 & b_1 & b_0
\end{matrix}\right) \left(\begin{matrix}
& \lambda a_l \\
& \lambda a_{l-1} \\
& \vdots \\
& \lambda a_2-a_4 \\
& \lambda a_1-a_2\\
& \lambda a_0-a_0
\end{matrix}\right)
\label{matrix2}
\end{equation}
where $b_j$, $j,0,...,l$ are the coefficients of $y^j$ in the expansion of
\begin{equation}
\frac1{a_l + a_{l-2} y + a_{l-4} y^2 +... + a^*_l y^l}=\frac1{y^{l/2}h_{odd}(y^{-1/2})}
\end{equation}
Since $P(z)+P(-z)=2$, we can fix $\lambda$ by requiring that $\sum_j^l p_{l-2j}=1$. A solution of Eq. \ref{matrix2} can only correspond to a valid choice of $P(z)$ if $p^{l-2j}$ thus found is indeed the complex conjugate of $p^{2j-l}$ for all $j$.

\bibliography{ehm,entanglement}



\end{document}